\documentclass[12pt]{article}
\usepackage{epsfig}
\newcommand{\mysection}{\setcounter{equation}{0}\section}

\def\beq{\begin{equation}}
\def\eeq{\end{equation}}
\def\beqa{\begin{eqnarray}}
\def\eeqa{\end{eqnarray}}

\newlength{\dinwidth} \newlength{\dinmargin}
\begin{document}
\begin {flushright}
Cavendish-HEP-03/16\\
LBNL-53599\\
\end {flushright} 
\vspace{3mm}
\begin{center}
{\Large \bf Next-to-next-to-leading order soft-gluon corrections 
in top quark hadroproduction}
\end{center}
\vspace{2mm}
\begin{center}
{\large Nikolaos Kidonakis$^a$ and Ramona Vogt$^b$}\\
\vspace{2mm}
$^a${\it Cavendish Laboratory, University of Cambridge,\\
Madingley Road, Cambridge CB3 0HE, UK}\\
\vspace{2.5mm}
$^b${\it Nuclear Science Division,\\
    Lawrence Berkeley National Laboratory, Berkeley, CA 94720, USA \\
  and \\
  Physics Department,\\
    University of California at Davis, Davis, CA 95616, USA}
\end{center}

\begin{abstract}
We calculate next-to-next-to-leading order soft-gluon corrections
to top quark total and differential cross sections in hadron colliders.
We increase the accuracy of our previous estimates by including
additional subleading terms, including 
next-to-next-to-next-to-leading-logarithmic and some virtual terms. 
We show that the kinematics dependence of the cross section vanishes
near threshold and is reduced away from it. 
The factorization and renormalization
scale dependence of the cross section is also greatly reduced.
We present results for  the top quark total cross sections
and transverse momentum distributions at the Tevatron and the LHC.
\end{abstract}

\thispagestyle{empty}  \newpage  \setcounter{page}{2}

\mysection{Introduction}

The discovery of the top quark in $p {\bar p}$ collisions
at Run I of the Tevatron in 1995 \cite{CDFD0} and its observation currently
at Run II, with expected increases in the accuracy of the top mass
and cross section measurements, have made theoretical calculations of 
top production cross sections and differential distributions
an interesting and topical subject.
The latest calculation for top hadroproduction 
includes next-to-next-to-leading-order (NNLO) soft-gluon 
corrections to the double differential cross section \cite{NKtop,KLMV} 
from threshold resummation techniques \cite{KS,KOSr,LOS}. 
Near threshold there is limited phase space for the emission of real
gluons so that soft-gluon corrections dominate the cross section.

These soft corrections take the form of logarithms,
$[\ln^l(x_{\rm th})/x_{\rm th}]_+$, with $l \le 2n-1$ for the 
order $\alpha_s^n$ corrections, where $x_{\rm th}$ is a kinematical
variable that measures distance from threshold and goes to zero
at threshold.
NNLO calculations for top quark production have so far been done
through next-to-next-to-leading-logarithmic (NNLL) accuracy, i.e.
including leading logarithms (LL) with $l=3$, 
next-to-leading logarithms (NLL) with $l=2$, and  
NNLL with $l=1$, at NNLO \cite{NKtop,KLMV}.
This NNLO-NNLL calculation has had great success in 
significantly reducing
the factorization/renormalization scale dependence of the cross
section.  Indeed the scale dependence of top production is almost negligible.
However, the dependence of the corrections on the kinematics choice
is substantial.
In Ref.~\cite{KLMV}, the top cross section was studied in both 
single-particle-inclusive (1PI) and pair-invariant-mass (PIM) kinematics.
Important differences  between the two kinematics choices
were found in both the parton-level and hadron-level cross sections, even
near threshold. Similar kinematics effects were found for bottom and charm 
hadroproduction \cite{KLMV,KLMVbc}. 
Thus subleading, beyond NNLL, contributions can still have
an impact on the cross section. If all the NNLO soft corrections 
are included,
there should be no difference between the two kinematics near threshold.
If all NNLO corrections, both soft and hard, were known, there should be no 
difference between the two kinematics, even far from threshold.
Away from threshold, where the approximations of Ref.~\cite{KLMV}
are not expected to apply since real emission of hard gluons comes into play,
the discrepancy between the 1PI and PIM results is not surprising. 
However, the NNLO-NNLL calculation exhibits some notable discrepancies
between the two kinematics even at the lowest $\eta$, where
$\eta=s/(4 m^2 )-1 \rightarrow 0$ at threshold. Thus, additional
subleading terms are clearly needed to bring the calculation
under further theoretical control.

In this paper, we include additional subleading NNLO soft corrections,
including next-to-next-to-next-to-leading logarithms (NNNLL) with $l=0$, 
as well as some virtual $\delta(x_{\rm th})$ corrections.
We apply the method and results of Ref.~\cite{NKNNLO}, based on earlier 
resummation studies \cite{KS,KOSr,LOS}, where master
formulas are given for the NNLO soft and virtual corrections for processes
in hadron-hadron and lepton-hadron collisions.
As we will see, the subleading corrections do indeed bring the 1PI and 
PIM results into agreement near threshold for both the
$q {\overline q} \rightarrow t {\overline t}$ and the
$gg \rightarrow t {\overline t}$ channels,
while the discrepancies away 
from threshold are also diminished, especially in the $gg$ channel.
Thus the threshold region is brought under theoretical control.
 
Since the resummation formalism has been reviewed extensively
in Refs.~\cite{NKtop,KLMV,KS,KOSr,NKNNLO}, we only provide a rough
outline here. Threshold resummation is a method of formally calculating 
contributions from soft-gluon emission to all
orders in perturbation theory.
The resummation is normally carried out in moment space where $N$ is the
variable conjugate to $x_{\rm th}$ and the leading threshold 
logarithms are of the form $\ln^{2n}N$ for the order 
$\alpha_s^n$ corrections.
The resummed cross section, in moment space, can then be expanded to 
NNLO (and even higher orders \cite{NKtop})  and the finite-order 
result finally inverted back to momentum space.
The previous calculations of Refs.~\cite{NKtop,KLMV} and the universal
results of Ref.~\cite{NKNNLO} employ this approach.

In the following section we give the analytical form of the NNLO soft and
(some) virtual corrections in the $q {\overline q} \rightarrow t \overline t$
channel in both 1PI and PIM kinematics. 
In Section 3 we give the results for the $gg \rightarrow t \overline t$ 
channel.  Note that while we refer only to $t \overline t$ here, the results in
Sections 2 and 3 are equally valid for all heavy quarks.
Section 4 discusses the partonic cross sections in both channels.
In Section 5 we present the hadronic cross sections and transverse
momentum distributions for top production in Tevatron Run I and Run II
as well as at the LHC. We conclude with a summary in Section 6.

\mysection{NNLO soft corrections to
$q {\overline q} \rightarrow t {\overline t}$}

We first study the $q {\overline q} \rightarrow t {\overline t}$ process. 
Before discussing the corrections, we introduce our kinematics notation 
(the same notation will also be used for $gg \rightarrow 
t {\overline t}$).
In 1PI kinematics the process is
\beq
q(p_a) +{\overline q}(p_b) \longrightarrow t(p_1) + X[{\overline t}](p_2)
\eeq
where $t$ is an identified top quark of mass $m$
and $X[{\overline t}]$ is the remaining final state that contains
the ${\overline t}$.
We define the kinematical invariants 
$s=(p_a+p_b)^2$, $t_1=(p_b-p_1)^2-m^2$, $u_1=(p_a-p_1)^2-m^2$
and $s_4=s+t_1+u_1$. At threshold, $s_4 \rightarrow 0$,
and the soft corrections appear as $[\ln^l(s_4/m^2)/s_4]_+$.

In PIM kinematics, we have instead
\beq
q(p_a) +{\overline q}(p_b) \longrightarrow t{\overline t}(p) + X(k) \, .
\label{PIM}
\eeq
At partonic threshold, $s=M^2$, $M^2$ is the pair mass squared, and
$t_1=-(M^2/2)(1-\beta_M \cos\theta)$, 
$u_1=-(M^2/2)(1+\beta_M \cos\theta)$
where $\beta_M=\sqrt{1-4m^2/M^2}$ and $\theta$ is the scattering
angle in the parton center-of-mass frame.
The soft corrections appear as $[\ln^l(1-z)/(1-z)]_+$
with $z=M^2/s \rightarrow 1$ at threshold.

A more detailed discussion of the kinematics can be found in Ref.~\cite{KLMV}.

\subsection{$q {\overline q} \rightarrow t {\overline t}$ channel in 
1PI kinematics}

We begin our study with the next-to-leading order (NLO) corrections.  
In the ${\overline {\rm MS}}$ scheme, the NLO soft and virtual corrections 
for $q {\overline q} \rightarrow t {\overline t}$
in 1PI kinematics can be written as
\beq
s^2\, \frac{d^2{\hat\sigma}^{(1)\; \rm 1PI}_{q {\overline q}}}{dt_1 \, du_1}
=F^{B\; \rm 1PI}_{q {\overline q}} 
\frac{\alpha_s(\mu_R^2)}{\pi} \left\{
c^{\rm 1PI}_{3 \;q {\overline q}} \left[\frac{\ln(s_4/m^2)}{s_4}\right]_+
+c^{\rm 1PI}_{2 \;q {\overline q}} \left[\frac{1}{s_4}\right]_+
+c^{\rm 1PI}_{1 \;q {\overline q}}  \, \delta(s_4) \right\}\, .
\label{NLO1PI}
\eeq
Here, the Born term is
\beq
F^{B\; \rm 1PI}_{q {\overline q}}=\pi\alpha_s^2(\mu_R^2)
K_{q\overline{q}} N_c C_F \left[ \frac{t_1^2 + u_1^2}{s^2} 
+ \frac{2m^2}{s}\right]\,
\eeq
where $\mu_R$ is the renormalization scale, $C_F=(N_c^2-1)/(2N_c)$
with $N_c=3$ the number of colors, and
$K_{q\overline{q}}=N_c^{-2}$ is a color average factor.

We also have
$c^{\rm 1PI}_{3 \;q {\overline q}}=4C_F$ and
\beqa
c^{\rm 1PI}_{2 \;q {\overline q}}&=&2C_F\left[4\ln\left(\frac{u_1}{t_1}\right)
-\ln\left(\frac{t_1u_1}{m^4}\right)-L'_{\beta}-1
-\ln\left(\frac{\mu_F^2}{s}\right)\right]
\nonumber \\ &&
{}+C_A\left[-3\ln\left(\frac{u_1}{t_1}\right)
-\ln\left(\frac{m^2s}{t_1u_1}\right)+L'_{\beta}\right] \, , 
\eeqa
where $\mu_F$ is the factorization scale, $C_A=N_c$,
$L'_{\beta}=[(1-2m^2/s)/\beta] \ln[(1-\beta)/(1+\beta)]$
and $\beta=\sqrt{1-4m^2/s}$.
For later use, we write 
\beq
c^{\rm 1PI}_{2 \;q {\overline q}} \equiv T^{\rm 1PI}_{2 \;q {\overline q}}
-2C_F\ln\left(\frac{\mu_F^2}{s}\right)\, ,
\eeq
so that $T^{\rm 1PI}_{2 \;q {\overline q}}$ is the scale-independent 
part of $c^{\rm 1PI}_{2 \;q {\overline q}}$.
Finally,
\beq
c^{\rm 1PI}_{1 \;q {\overline q}}=
\frac{\sigma^{(1)\, S+V\; \rm 1PI}_{q {\overline q} \, \delta}}
{(\alpha_s/\pi) F^{B\; \rm 1PI}_{q {\overline q}}} \, 
\label{c1}
\eeq
where $\sigma^{(1)\, S+V\; \rm 1PI}_{q {\overline q} \, \delta}$ 
denotes the $\delta(s_4)$ terms
in Eq. (4.7) of Ref. \cite{BNMSJ} with the definitions of $t_1$ and $u_1$ 
interchanged with respect to that reference.
We also write
\beq
c^{\rm 1PI}_{1 \;q {\overline q}} \equiv T^{\rm 1PI}_{1 \;q {\overline q}}
+C_F \left[-\frac{3}{2}+\ln\left(\frac{t_1u_1}{m^4}\right)\right]
\ln\left(\frac{\mu_F^2}{s}\right) 
+\frac{\beta_0}{2}\ln\left(\frac{\mu_R^2}{s}\right) \, ,
\eeq
where $T^{\rm 1PI}_{1 \;q {\overline q}}$ has no scale dependence.

Before presenting the NNLO soft corrections, 
we define the constants 
$\zeta_2=\pi^2/6$, $\zeta_4=\pi^4/90$, and $\zeta_3=1.2020569\cdots$,
and the two-loop constant $K=C_A(67/18-\pi^2/6)-5n_f/9$, with $n_f$
the number of light quark flavors.
Finally, we define
\beq
\beta(\alpha_s) \equiv \mu \, \frac{d \ln g}{d \mu}
=-\beta_0 \frac{\alpha_s}{4 \pi}-\beta_1 \frac{\alpha_s^2}{(4 \pi)^2} +\cdots,
\eeq 
where $\beta_0=(11C_A-2n_f)/3$ and
\beq
\beta_1=\frac{34}{3} C_A^2-2 n_f\left(C_F+\frac{5}{3}C_A\right) \, .
\eeq

Following Ref. \cite{NKNNLO} we write the NNLO soft-plus-virtual 
corrections in 1PI kinematics as
\beqa
&& \hspace{-8mm}s^2\, \frac{d^2{\hat\sigma}^{(2)\; \rm 1PI}_{q {\overline q}}}
{dt_1 \, du_1}
=F^{B\; \rm 1PI}_{q {\overline q}} \frac{\alpha_s^2(\mu_R^2)}{\pi^2} 
\left\{\frac{1}{2} \left(c^{\rm 1PI}_{3 \;q {\overline q}}\right)^2 
\left[\frac{\ln^3(s_4/m^2)}{s_4}\right]_+
+\left[\frac{3}{2} c^{\rm 1PI}_{3 \;q {\overline q}} \, c^{\rm 1PI}_{2 \;q {\overline q}}
-\frac{\beta_0}{4} c_{3 \;q {\overline q}} ^{\rm 1PI} \right] 
\left[\frac{\ln^2(s_4/m^2)}{s_4}\right]_+
\right.
\nonumber \\ && 
{}+\left[c^{\rm 1PI}_{3 \;q {\overline q}} \, c^{\rm 1PI}_{1 \;q {\overline q}}
+\left(c^{\rm 1PI}_{2 \;q {\overline q}}\right)^2
-\zeta_2 \, \left(c^{\rm 1PI}_{3 \;q {\overline q}}\right)^2
-\frac{\beta_0}{2} T^{\rm 1PI}_{2 \;q {\overline q}}
+\frac{\beta_0}{4} c_{3\;q {\overline q}}^{\rm 1PI} 
\ln\left(\frac{\mu_R^2}{s}\right) \right.
\nonumber \\ && \quad \quad \left.
{}+2 C_F K+8 \frac{C_F}{C_A}\ln^2\left(\frac{u_1}{t_1}\right)\right] 
\left[\frac{\ln(s_4/m^2)}{s_4}\right]_+
\nonumber \\ && 
{}+\left[c^{\rm 1PI}_{2 \;q {\overline q}} \, c^{\rm 1PI}_{1 \;q {\overline q}}
-\zeta_2 \, c^{\rm 1PI}_{2 \;q {\overline q}} \, c^{\rm 1PI}_{3 \;q {\overline q}}
+\zeta_3 \, \left(c^{\rm 1PI}_{3 \;q {\overline q}}\right)^2 
-\frac{\beta_0}{2} T_{1\;q {\overline q}}^{\rm 1PI} 
+\frac{\beta_0}{4} c_{2\;q {\overline q}}^{\rm 1PI} \ln\left(\frac{\mu_R^2}{s}\right)
+{\cal G}_{q{\overline q}}^{(2)}\right.
\nonumber \\ && \quad \quad \left.
{}+C_F\frac{\beta_0}{4} \ln^2\left(\frac{\mu_F^2}{s}\right)
-C_F K \ln\left(\frac{\mu_F^2}{s}\right)
-C_F K \ln\left(\frac{t_1 u_1}{m^4}\right)
+8 \frac{C_F}{C_A}\ln^2\left(\frac{u_1}{t_1}\right)
\ln\left(\frac{m^2}{s}\right) \right]
\left[\frac{1}{s_4}\right]_+
\nonumber \\ &&   \left.  
{}+R^{\rm 1PI}_{q {\overline q}} \delta(s_4)\right\} \, .
\label{NNLOqqMS}
\eeqa
Here
\beq
{\cal G}_{q \overline q}^{(2)}=C_F C_A 
\left(\frac{7}{2} \zeta_3
+\frac{22}{3}\zeta_2-\frac{299}{27}\right)+ n_f C_F \left(-\frac{4}{3}\zeta_2
+\frac{50}{27}\right) 
\eeq
denotes a set of two-loop contributions that are universal for processes
with $q {\overline q}$ initial states \cite{NKNNLO}.  Process-dependent
two-loop corrections \cite{NK2l} are not included in 
${\cal G}_{q \overline q}^{(2)}$ but, as we will see in Section 4, their
contribution is expected to be negligible.
The virtual contribution $R^{\rm 1PI}_{q {\overline q}}$ is not fully known. 
However, we can determine certain terms in $R^{\rm 1PI}_{q {\overline q}}$
exactly.  These exact terms involve the factorization
and renormalization scales as well as the those terms that arise from the
inversion from moment to momentum space (for a detailed discussion
of the inversion procedure see Section IIIC and Appendix A of 
Ref. \cite{NKtop}).

The terms multiplying $\delta(s_4)$ involving
the factorization and renormalization scales are given explicitly by
\beqa
&& \hspace{-10mm} F^{B\; \rm 1PI}_{q {\overline q}}
\frac{\alpha_s^2(\mu_R^2)}{\pi^2} \left[ \ln^2\left(\frac{\mu_F^2}{m^2}\right)
\left\{\frac{C_F^2}{2}\left[\ln\left(\frac{t_1 \, u_1}{m^4}\right)
-\frac{3}{2}\right]^2-2 \zeta_2 C_F^2
+ \frac{\beta_0}{8} C_F \left[\frac{3}{2} 
-\ln\left(\frac{t_1 \, u_1}{m^4}\right)\right]\right\} \right.
\nonumber \\ &&
{}+\ln\left(\frac{\mu_F^2}{m^2}\right)
\ln\left(\frac{\mu_R^2}{m^2}\right)
\frac{3\beta_0}{4}C_F\left[\ln\left(\frac{t_1 \, u_1}{m^4}\right)
-\frac{3}{2}\right]
+\ln^2 \left(\frac{\mu_R^2}{m^2}\right) \, \frac{3 \beta_0^2}{16}
\nonumber \\ &&
{}+\ln\left(\frac{\mu_F^2}{m^2}\right)
\left\{C_F^2 \left[\ln\left(\frac{t_1 \, u_1}{m^4}\right)
-\frac{3}{2}\right]^2 \ln\left(\frac{m^2}{s}\right)
+C_F \left[\ln\left(\frac{t_1 \, u_1}{m^4}\right)-\frac{3}{2}\right] 
\left[T^{\rm 1PI}_{1 \; q{\overline q}}
+\frac{\beta_0}{2} \ln\left(\frac{m^2}{s}\right)\right] \right.
\nonumber \\ && \hspace{25mm} \left.
{}+2 C_F \zeta_2 \left[T^{\rm 1PI}_{2 \; q{\overline q}}-2C_F
\ln\left(\frac{m^2}{s}\right)\right] -8 C_F^2 \zeta_3
+C_F \frac{K}{2} \ln\left(\frac{t_1 \, u_1}{m^4}\right)
-2 {\gamma'}^{(2)}_{q/q} \right\}
\nonumber \\ && \left.
{}+\ln\left(\frac{\mu_R^2}{m^2}\right) \left\{\frac{3\beta_0}{4}
\left[C_F\left(\ln\left(\frac{t_1 \, u_1}{m^4}\right)
-\frac{3}{2}\right)\ln\left(\frac{m^2}{s}\right)
+\frac{\beta_0}{2} \ln\left(\frac{m^2}{s}\right)
+T^{\rm 1PI}_{1 \; q{\overline q}}\right]
+\frac{\beta_1}{8}\right\} \right]
\label{Rmu}
\eeqa
where
\beq
{\gamma'}_{q/q}^{(2)}=C_F^2\left(\frac{3}{32}-\frac{3}{4}\zeta_2
+\frac{3}{2}\zeta_3\right)
+C_F C_A\left(-\frac{3}{4}\zeta_3+\frac{11}{12}\zeta_2+\frac{17}{96}\right)
+n_f C_F \left(-\frac{\zeta_2}{6}-\frac{1}{48}\right)\, .
\eeq

The terms multiplying $\delta(s_4)$ resulting from inversion ($\zeta$ terms)
that do not involve the factorization and renormalization scales are
\cite{NKtop,NKNNLO}
\beqa
&& \hspace{-5mm}  F^{B\; \rm 1PI}_{q {\overline q}}
\frac{\alpha_s^2(\mu_R^2)}{\pi^2} \left\{
-\frac{\zeta_2}{2}\, \left[T^{\rm 1PI}_{2 \;q {\overline q}}
-2C_F \ln\left(\frac{m^2}{s}\right)\right]^2 
+\frac{1}{4}\zeta_2^2 \,\left(c^{\rm 1PI}_{3 \;q {\overline q}}\right)^2
+\zeta_3 \, c^{\rm 1PI}_{3 \;q {\overline q}} \, 
\left[T^{\rm 1PI}_{2 \;q {\overline q}}
-2C_F \ln\left(\frac{m^2}{s}\right)\right] \right.
\nonumber \\ && \hspace{22mm} \left.
{}-\frac{3}{4}\zeta_4 \, \left(c^{\rm 1PI}_{3 \;q {\overline q}}\right)^2 
-4 \zeta_2 \frac{C_F}{C_A}\ln^2\left(\frac{u_1}{t_1}\right) \right\} \, .
\eeqa

\subsection{$q {\overline q} \rightarrow t {\overline t}$ 
channel in PIM kinematics}

Next, we study the soft-gluon corrections in PIM kinematics.
The ${\overline {\rm MS}}$ NLO soft and virtual corrections to 
$q {\overline q} \rightarrow t {\overline t}$ in PIM kinematics are
\beq
s\, \frac{d^2{\hat\sigma}^{(1)\; {\rm PIM}}_{q {\overline q}}}{dM^2 \, d\cos\theta}
=F^{B\; {\rm PIM}}_{q {\overline q}} 
\frac{\alpha_s(\mu_R^2)}{\pi} 
\left\{c^{{\rm PIM}}_{3 \;q {\overline q}} \left[\frac{\ln(1-z)}{1-z}\right]_+
+c^{{\rm PIM}}_{2 \;q {\overline q}} \left[\frac{1}{1-z}\right]_+
+c^{{\rm PIM}}_{1 \;q {\overline q}}  \delta(1-z) \right\}\, .
\label{NLOPIM}
\eeq
Here the Born term is
\beq
F^{B\; {\rm PIM}}_{q {\overline q}}=\frac{\beta}{2s}F^{B\; \rm 1PI}_{q {\overline q}}|_{{\rm PIM}}
=\frac{\beta}{2s} \pi\alpha_s^2 
K_{q\overline{q}} N_c C_F \left[\frac{1}{2}\left(1+\beta^2 \cos^2\theta\right) 
+ \frac{2m^2}{s}\right]\, ,
\eeq
where $|_{\rm PIM}$ indicates that for $t_1$, $u_1$ we use the expressions
below Eq. (\ref{PIM}).
Also $c^{{\rm PIM}}_{3 \;q {\overline q}}=4C_F$,
\beqa
c^{\rm PIM}_{2 \;q {\overline q}}&=&
2C_F\left[4\ln\left(\frac{u_1}{t_1}\right)
-L'_{\beta}-1
-\ln\left(\frac{\mu_F^2}{s}\right)\right]
\nonumber \\ &&
{}+C_A\left[-3\ln\left(\frac{u_1}{t_1}\right)
-\ln\left(\frac{m^2s}{t_1u_1}\right)+L'_{\beta}\right] \, , 
\nonumber \\ 
&\equiv& T^{{\rm PIM}}_{2 \;q {\overline q}}
- 2C_F\ln\left(\frac{\mu_F^2}{s}\right) \, ,
\eeqa
and
\beq
c^{{\rm PIM}}_{1 \;q {\overline q}} \equiv T^{{\rm PIM}}_{1 \;q {\overline q}}
-\frac{3}{2} C_F \ln\left(\frac{\mu_F^2}{s}\right) 
+\frac{\beta_0}{2}\ln\left(\frac{\mu_R^2}{s}\right) \, .
\eeq
Note that the scale-independent $T^{{\rm PIM}}_{1 \;q {\overline q}}$
is related to its 1PI counterpart by
\beq
T^{{\rm PIM}}_{1 \;q {\overline q}}=2 
T^{\rm 1PI}_{1 \;q {\overline q}}|_{{\rm PIM}}
+\frac{1}{F^{B\; {\rm PIM}}_{q {\overline q}}} \;
s\, \frac{d^2{\sigma'}^{(1)\; {\rm S+MF}}_{q {\overline q}}}{dM^2 \,
d\cos\theta}
-\frac{1}{F^{B\; {\rm PIM}}_{q {\overline q}}} \; \frac{\beta}{s} \;
s^2\, \frac{d^2{\sigma'}^{(1)\; {\rm S+MF}}_{q {\overline q}}}{dt_1 \, 
du_1}|_{{\rm PIM}} \, .
\eeq
Here ${\sigma'}^{(1)\; {\rm S+MF}}_{q {\overline q}}$ 
denotes the soft and mass factorization 
subtraction terms calculated in Ref.~\cite{KLMV}.
The prime indicates that we drop the overall $\delta(1-z)$ or
$\delta(s_4)$ coefficient from the expressions in Eqs.~(82), (A8), and (A9)
of Ref.~\cite{KLMV}.

In PIM kinematics, the NNLO soft-plus-virtual corrections are
\beqa
&& \hspace{-10mm} 
s\, \frac{d^2{\hat\sigma}^{(2)\; {\rm PIM}}_{q {\overline 
q}}}{dM^2 \, d\cos\theta}
=F^{B\; {\rm PIM}}_{q {\overline q}} \frac{\alpha_s^2(\mu_R^2)}{\pi^2} 
\left\{\frac{1}{2} \left(c^{{\rm PIM}}_{3 \;q {\overline q}}\right)^2 
\left[\frac{\ln^3(1-z)}{1-z}\right]_+
+\left[\frac{3}{2} c^{{\rm PIM}}_{3 \;q {\overline q}} \, 
c^{{\rm PIM}}_{2 \;q {\overline q}}
-\frac{\beta_0}{4} c_{3 \;q {\overline q}} ^{{\rm PIM}} \right] 
\left[\frac{\ln^2(1-z)}{1-z}\right]_+
\right.
\nonumber \\ && 
{}+\left[c^{{\rm PIM}}_{3 \;q {\overline q}} \, 
c^{{\rm PIM}}_{1 \;q {\overline q}}
+\left(c^{{\rm PIM}}_{2 \;q {\overline q}}\right)^2
-\zeta_2 \, \left(c^{{\rm PIM}}_{3 \;q {\overline q}}\right)^2
-\frac{\beta_0}{2} T^{{\rm PIM}}_{2 \;q {\overline q}}
+\frac{\beta_0}{4} c_{3\;q {\overline q}}^{{\rm PIM}} 
\ln\left(\frac{\mu_R^2}{s}\right) \right.
\nonumber \\ && \quad \quad \left.
{}+2 C_F K+8 \frac{C_F}{C_A}\ln^2\left(\frac{u_1}{t_1}\right) \right] 
\left[\frac{\ln(1-z)}{1-z}\right]_+
\nonumber \\ && 
{}+\left[c^{{\rm PIM}}_{2 \;q {\overline q}} \, c^{{\rm PIM}}_{1 \;q {\overline q}}
-\zeta_2 \, c^{{\rm PIM}}_{2 \;q {\overline q}} \, c^{{\rm PIM}}_{3 \;q {\overline q}}
+\zeta_3 \, \left(c^{{\rm PIM}}_{3 \;q {\overline q}}\right)^2 
-\frac{\beta_0}{2} T_{1\;q {\overline q}}^{{\rm PIM}} 
+\frac{\beta_0}{4} c_{2\;q {\overline q}}^{{\rm PIM}} \ln\left(\frac{\mu_R^2}{s}\right)
+{\cal G}_{q{\overline q}}^{(2)}\right.
\nonumber \\ && \quad \quad \left.
{}+C_F\frac{\beta_0}{4} \ln^2\left(\frac{\mu_F^2}{s}\right)
-C_F K \ln\left(\frac{\mu_F^2}{s}\right) \right]
\left[\frac{1}{1-z}\right]_+
\nonumber \\ && \left. 
+R^{{\rm PIM}}_{q{\overline q}}  \delta(1-z)\right\}\, .
\label{NNLOqqPIMMS}
\eeqa

Again, only certain terms in $R^{{\rm PIM}}_{q{\overline q}}$ that can be 
determined exactly are included.
The terms multiplying  $\delta(1-z)$ that involve
the factorization and renormalization scales are 
\beqa
&& \hspace{-15mm} F^{B\; {\rm PIM}}_{q {\overline q}}
\frac{\alpha_s^2(\mu_R^2)}{\pi^2} \left[ \ln^2\left(\frac{\mu_F^2}{m^2}\right)
\left\{\frac{9}{8} C_F^2 -2 \zeta_2 C_F^2
+\frac{3}{16} C_F \beta_0 \right\} \right.
\nonumber \\ &&
{}-\frac{9}{8} \beta_0 \, C_F \ln\left(\frac{\mu_F^2}{m^2}\right)
\ln\left(\frac{\mu_R^2}{m^2}\right)
+\frac{3 \beta_0^2}{16} \, \ln^2 \left(\frac{\mu_R^2}{m^2}\right) 
\nonumber \\ &&
{}+\ln\left(\frac{\mu_F^2}{m^2}\right)
\left\{\frac{9}{4} C_F^2 \ln\left(\frac{m^2}{s}\right)
-\frac{3}{2} C_F \left[T^{{\rm PIM}}_{1 \; q{\overline q}}
+\frac{\beta_0}{2} \ln\left(\frac{m^2}{s}\right)\right] \right.
\nonumber \\ && \hspace{25mm} \left.
{}+2 C_F \zeta_2 \left[T^{{\rm PIM}}_{2 \; q{\overline q}}-2C_F
\ln\left(\frac{m^2}{s}\right)\right] -8 C_F^2 \zeta_3
-2 {\gamma'}^{(2)}_{q/q} \right\}
\nonumber \\ && \left.
{}+\ln\left(\frac{\mu_R^2}{m^2}\right) \left\{\frac{3\beta_0}{4}
\left[-\frac{3}{2} C_F \ln\left(\frac{m^2}{s}\right)
+\frac{\beta_0}{2} \ln\left(\frac{m^2}{s}\right)
+T^{{\rm PIM}}_{1 \; q{\overline q}}\right]
+\frac{\beta_1}{8}\right\} \right] \, .
\label{RmuPIM}
\eeqa

The terms multiplying $\delta(1-z)$ that arise from inversion and do not 
involve the factorization and renormalization scales are given by
\beqa
&& \hspace{-20mm} F^{B\; {\rm PIM}}_{q {\overline q}}
\frac{\alpha_s^2(\mu_R^2)}{\pi^2} \left\{
-\frac{\zeta_2}{2}\, \left[T^{{\rm PIM}}_{2 \;q {\overline q}}
-2C_F \ln\left(\frac{m^2}{s}\right)\right]^2
+\frac{1}{4}\zeta_2^2 \,\left(c^{{\rm PIM}}_{3 \;q {\overline q}}\right)^2
\right.
\nonumber \\ && \quad \quad \left.
{}+\zeta_3 \, c^{{\rm PIM}}_{3 \;q {\overline q}} \, \left[T^{{\rm PIM}}_{2 
\;q {\overline q}}
-2C_F \ln\left(\frac{m^2}{s}\right)\right]
-\frac{3}{4}\zeta_4 \, \left(c^{{\rm PIM}}_{3 \;q {\overline q}}\right)^2 
-4 \zeta_2 \frac{C_F}{C_A}\ln^2\left(\frac{u_1}{t_1}\right) \right\} \, .
\eeqa

\mysection{NNLO soft corrections to $gg \rightarrow t 
{\overline t}$}

\subsection{$gg \rightarrow t {\overline t}$ channel in \rm 1PI kinematics}

We now turn to the $gg$ channel. We write the 
${\overline {\rm MS}}$ NLO soft-plus-virtual
corrections for $gg \rightarrow  t{\overline t}$
in 1PI kinematics as
\beqa
s^2\, \frac{d^2{\hat\sigma}^{(1)\; \rm 1PI}_{gg}}{dt_1 \, du_1}
&=&F^{B\; \rm 1PI}_{gg} 
\frac{\alpha_s(\mu_R^2)}{\pi} 
\left\{c^{\rm 1PI}_{3 \; gg} \left[\frac{\ln(s_4/m^2)}{s_4}\right]_+
+c^{\rm 1PI}_{2 \; gg} \left[\frac{1}{s_4}\right]_+
+c^{\rm 1PI}_{1 \; gg}  \delta(s_4)\right\}
\nonumber \\ &&
{}+\frac{\alpha_s^{3}(\mu_R^2)}{\pi} 
\left[A^c_{gg} \,\left[\frac{1}{s_4}\right]_+ 
+T^{c\, \rm 1PI}_{1\; gg} \, \delta(s_4)\right]\, .
\label{NLOgg1PI}
\eeqa
The Born term is given by 
\beq
F^{B\; \rm 1PI}_{gg} =  2\pi \alpha_s^2(\mu_R^2) K_{gg}
N_c C_F \left[C_F - C_A \frac{t_1u_1}{s^2}\right] B_{\rm QED} \, ,
\eeq
where $K_{gg}=(N_c^2-1)^{-2}$ is a color average factor  and
\beq
B_{\rm QED}=\frac{t_1}{u_1}+\frac{u_1}{t_1}+\frac{4m^2s}{t_1u_1}
\left(1-\frac{m^2s}{t_1u_1}\right) \, .
\eeq
We also define
$c^{\rm 1PI}_{3 \; gg}=4C_A$,
\beq
c^{\rm 1PI}_{2 \; gg}=-2C_A-2C_A\ln\left(\frac{t_1u_1}{m^4}\right)
-2C_A\ln\left(\frac{\mu_F^2}{s}\right) \equiv T^{\rm 1PI}_{2 \; gg}
-2C_A\ln\left(\frac{\mu_F^2}{s}\right)\, ,
\eeq
\beq
c^{\rm 1PI}_{1 \; gg}=\left[C_A \ln\left(\frac{t_1u_1}{m^4}\right)
-\frac{\beta_0}{2}\right]
\ln\left(\frac{\mu_F^2}{s}\right) 
+\frac{\beta_0}{2}\ln\left(\frac{\mu_R^2}{s}\right)\, ,
\eeq
and
\beqa
A^c_{gg}&=& \pi K_{gg} B_{\rm QED} (N_c^2-1)\left\{N_c
\left(1-\frac{2t_1u_1}{s^2}\right)
\left[\left(-C_F+\frac{C_A}{2}\right)({\rm Re}L_{\beta}+1)
+\frac{N_c}{2}+\frac{N_c}{2} \ln\left(\frac{t_1u_1}{m^2 s}\right)\right]
\right.
\nonumber \\ && \left.
{}+\frac{1}{N_c}(C_F-C_A)({\rm Re}L_{\beta}+1)
-\ln\left(\frac{t_1u_1}{m^2 s}\right)
+\frac{N_c^2}{2} \frac{(t_1^2-u_1^2)}{s^2} \ln\left(\frac{u_1}{t_1}\right)
\right\}\, . \label{acgg}
\eeqa
Finally,
\beq
T^{c\, \rm 1PI}_{1\; gg}=\frac{{\sigma}^{(1)\, {\rm S+V} \, \rm 1PI}_{gg \, \delta}}
{\alpha_s^3/\pi}
\eeq
where here $\sigma^{(1)\, {\rm S+V} \, \rm 1PI}_{gg \, \delta}$ 
denotes the scale-independent $\delta(s_4)$ terms in 
the NLO cross section. 
These terms are given by Eq.~(6.19) in Ref.~\cite{NLOgg}.

The NNLO soft-plus-virtual corrections in 1PI kinematics are
\beqa
&& \hspace{-10mm} s^2\, \frac{d^2{\hat\sigma}^{(2)\; 
\rm 1PI}_{gg}}{dt_1 \, du_1}
=F^{B\; \rm 1PI}_{gg} \frac{\alpha_s^2(\mu_R^2)}{\pi^2} 
\left\{\frac{1}{2} \left(c^{\rm 1PI}_{3 \;gg}\right)^2 
\left[\frac{\ln^3(s_4/m^2)}{s_4}\right]_+
+\left[\frac{3}{2} c^{\rm 1PI}_{3 \;gg} \, c^{\rm 1PI}_{2 \;q {\overline q}}
-\frac{\beta_0}{4} c_{3 \;gg} ^{\rm 1PI} \right] 
\left[\frac{\ln^2(s_4/m^2)}{s_4}\right]_+
\right.
\nonumber \\ && \hspace{-5mm}
{}+\left[c^{\rm 1PI}_{3 \;gg} \, c^{\rm 1PI}_{1 \;gg}
+\left(c^{\rm 1PI}_{2 \;gg}\right)^2
-\zeta_2 \, \left(c^{\rm 1PI}_{3 \;gg}\right)^2
-\frac{\beta_0}{2} T^{\rm 1PI}_{2 \;gg}
+\frac{\beta_0}{4} c_{3\;gg}^{\rm 1PI} \ln\left(\frac{\mu_R^2}{s}\right)
+2 C_A K\right]
\left[\frac{\ln(s_4/m^2)}{s_4}\right]_+
\nonumber \\ && \hspace{-5mm}
{}+\left[c^{\rm 1PI}_{2 \;gg} \, c^{\rm 1PI}_{1 \;gg}
-\zeta_2 \, c^{\rm 1PI}_{2 \;gg} \, c^{\rm 1PI}_{3 \;gg}
+\zeta_3 \, \left(c^{\rm 1PI}_{3 \;gg}\right)^2 
+\frac{\beta_0}{4} c_{2\;gg}^{\rm 1PI} \ln\left(\frac{\mu_R^2}{s}\right)
+{\cal G}_{gg}^{(2)}\right.
\nonumber \\ && \quad \left.
{}+C_A\frac{\beta_0}{4} \ln^2\left(\frac{\mu_F^2}{s}\right)
-C_A K \ln\left(\frac{\mu_F^2}{s}\right)
-C_A K \ln\left(\frac{t_1 u_1}{m^4}\right)\right]
\left[\frac{1}{s_4}\right]_+
\nonumber \\ && \hspace{-5mm} \left.
{}+R^{\rm 1PI}_{gg}  
\delta(s_4)\right\} 
\nonumber \\ && \hspace{-10mm}
{}+\frac{\alpha_s^{4}(\mu_R^2)}{\pi^2} 
\left\{\frac{3}{2} c^{\rm 1PI}_{3 \;gg} \, A^c_{gg}\, 
\left[\frac{\ln^2(s_4/m^2)}{s_4}\right]_+
+\left[\left(2\, c^{\rm 1PI}_{2 \;gg}-\frac{\beta_0}{2}\right)A^c_{gg}
+c^{\rm 1PI}_{3 \;gg} T^{c \, \rm 1PI}_{1 \;gg}
+F^c_{gg}\right] \left[\frac{\ln(s_4/m^2)}{s_4}\right]_+
\right.
\nonumber \\ && \hspace{-5mm}
{}+\left[\left(c^{\rm 1PI}_{1 \;gg}-\zeta_2 c^{\rm 1PI}_{3 \;gg}+\frac{\beta_0}{4}
\ln\left(\frac{\mu_R^2}{s}\right)\right)A^c_{gg}+\left(c^{\rm 1PI}_{2 \;gg}
-\frac{\beta_0}{2}\right)
T^{c \, \rm 1PI}_{1 \;gg}+F^c_{gg} \, \ln\left(\frac{m^2}{s}\right)\right] 
\left[\frac{1}{s_4}\right]_+
\nonumber \\ && \hspace{-5mm} \left. 
{}+ R^{c\, \rm 1PI}_{gg} \delta(s_4)\right\} \, ,
\label{NNLOgg}
\eeqa
where
\beqa
F^c_{gg}&=&\frac{\pi}{2} K_{gg} B_{\rm QED} (N_c^2-1) 
\left\{ 2\ln\left(\frac{u_1}{t_1}\right) 
\frac{(t_1^2-u_1^2)}{s^2} \left[4\, {\Gamma}^{gg}_{11}
+2(N_c^2-2) {\Gamma}^{gg}_{22}\right]\right.
\nonumber \\ &&
{}+\left(1-\frac{2t_1u_1}{s^2}\right)N_c\left[4
({\Gamma}^{gg}_{22})^2+(N_c^2+4)\ln^2\left(\frac{u_1}{t_1}\right) \right]
\nonumber \\ && \left.
{}+\frac{4}{N_c}\left[({\Gamma}^{gg}_{11})^2
-2 ({\Gamma}^{gg}_{22})^2\right]
-2N_c\ln^2\left(\frac{u_1}{t_1}\right)\right\}\, ,
\eeqa
with
\beqa
{\Gamma}^{gg}_{11}& \equiv & -C_F \left(L'_{\beta}+1\right)+C_A \, ,
\nonumber \\ 
{\Gamma}^{gg}_{22} & \equiv &-C_F \left(L'_{\beta}+1\right)
+\frac{C_A}{2}\left[2+\ln\left(\frac{t_1 u_1}{m^2 s}\right)
+L'_{\beta}\right] \, .
\eeqa
Here
\beq
{\cal G}_{gg}^{(2)}=C_A^2 \left(\frac{7}{2} \zeta_3
+\frac{22}{3}\zeta_2-\frac{41}{108}\right)+n_f C_A \left(-\frac{4}{3}\zeta_2
-\frac{5}{54}\right)\, 
\eeq
denotes a set of universal two-loop contributions for processes
with $gg$ initial states \cite{NKNNLO};  process-dependent
two-loop corrections \cite{NK2l} are not included in ${\cal G}_{gg}^{(2)}$. 

The contributions $R^{\rm 1PI}_{gg}$ and $R^{c\, \rm 1PI}_{gg}$ are 
virtual corrections that are not fully known.  As in the $q \overline q$
channel, we keep only certain terms that can be determined exactly.
The terms multiplying $\delta(s_4)$ involving
the factorization and renormalization scales are
\beqa
&& \hspace{-15mm} F^{B\; \rm 1PI}_{gg} \frac{\alpha_s^2(\mu_R^2)}{\pi^2} 
\left[\ln^2\left(\frac{\mu_F^2}{m^2}\right)
\left\{\frac{C_A^2}{2}\ln^2\left(\frac{t_1 \, u_1}{m^4}\right)
-\frac{5\beta_0}{8}C_A\ln\left(\frac{t_1 \, u_1}{m^4}\right) 
+\frac{3\beta_0^2}{16}-2 \zeta_2 C_A^2 \right\} \right.
\nonumber \\ &&
{}+\ln\left(\frac{\mu_F^2}{m^2}\right)
\ln\left(\frac{\mu_R^2}{m^2}\right)
\frac{3\beta_0}{4}\left[C_A\ln\left(\frac{t_1 \, u_1}{m^4}\right)
-\frac{\beta_0}{2}\right]
+\ln^2 \left(\frac{\mu_R^2}{m^2}\right) \, \frac{3 \beta_0^2}{16}
\nonumber \\ &&
{}+\ln\left(\frac{\mu_F^2}{m^2}\right)
\left\{C_A^2 \ln^2\left(\frac{t_1 \, u_1}{m^4}\right)
\ln\left(\frac{m^2}{s}\right)-\frac{\beta_0}{2}C_A
\ln\left(\frac{t_1 \, u_1}{m^4}\right)\ln\left(\frac{m^2}{s}\right) \right.
\nonumber \\ && \left. \hspace{20mm}
{}+2 C_A \zeta_2 \left[T^{\rm 1PI}_{2 \; gg}-2C_A
\ln\left(\frac{m^2}{s}\right)\right] -8 C_A^2 \zeta_3
+C_A \frac{K}{2} \ln\left(\frac{t_1 \, u_1}{m^4}\right)
-2 {\gamma'}^{(2)}_{g/g} \right\}
\nonumber \\ && \left.
{}+\ln\left(\frac{\mu_R^2}{m^2}\right) \left\{\frac{3\beta_0}{4}
C_A\ln\left(\frac{t_1 \, u_1}{m^4}\right)
\ln\left(\frac{m^2}{s}\right)
+\frac{\beta_1}{8}\right\}  \right]
\nonumber \\ && \hspace{-15mm}
{}+ \frac{\alpha_s^4(\mu_R^2)}{\pi^2}\left\{\left[2C_A \zeta_2 A^c_{gg}
+\left(C_A \ln\left(\frac{t_1 \, u_1}{m^4}\right)
-\frac{\beta_0}{2}\right)T^{c \, \rm 1PI}_{1 \; gg}\right] 
\ln\left(\frac{\mu_F^2}{m^2}\right)
+\frac{3\beta_0}{4} T^{c \, \rm 1PI}_{1 \; gg} \ln\left(\frac{\mu_R^2}{m^2}
\right)\right\}
\eeqa
where
\beq
{\gamma'}_{g/g}^{(2)}=C_A^2\left(\frac{2}{3}+\frac{3}{4}\zeta_3\right)
-n_f\left(\frac{C_F}{8}+\frac{C_A}{6}\right) \, . 
\eeq

The terms multiplying $\delta(s_4)$ that arise from inversion and do not 
involve
the factorization and renormalization scales are 
\beqa
&& \hspace{-20mm} F^{B\; \rm 1PI}_{gg} \frac{\alpha_s^2(\mu_R^2)}{\pi^2} 
\left\{-\frac{\zeta_2}{2}\, \left[T^{\rm 1PI}_{2 \;gg}
-2C_A \ln\left(\frac{m^2}{s}\right)\right]^2 
+\frac{1}{4}\zeta_2^2 \,\left(c^{\rm 1PI}_{3 \;gg}\right)^2 \right.
\nonumber \\ && \quad \quad \left.
+\zeta_3 \, c^{\rm 1PI}_{3 \;gg} \, \left[T^{\rm 1PI}_{2 \;gg}
-2C_A \ln\left(\frac{m^2}{s}\right)\right]
-\frac{3}{4}\zeta_4 \, \left(c^{\rm 1PI}_{3 \;gg}\right)^2  \right\}
\nonumber \\ && \hspace{-20mm}
{}+\frac{\alpha_s^4(\mu_R^2)}{\pi^2}
\left\{\left[\zeta_3 c^{\rm 1PI}_{3 \;gg}-\zeta_2 
\left(T^{\rm 1PI}_{2 \;gg}-
2C_A\ln\left(\frac{m^2}{s}\right)\right)\right]A^c_{gg}
-\frac{\zeta_2}{2}F^c_{gg} \right\} \, .
\eeqa

\subsection{$gg \rightarrow t {\overline t}$ channel in PIM kinematics}

We continue our study of subleading terms in the $gg$ channel by writing the 
${\overline {\rm MS}}$ NLO soft-plus-virtual
corrections for $gg \rightarrow t {\overline t}$
in PIM kinematics as
\beqa
s\, \frac{d^2{\hat\sigma}^{(1)\; {\rm PIM}}_{gg}}{dM^2 \, d\cos\theta}
&=&F^{B\; {\rm PIM}}_{gg} 
\frac{\alpha_s(\mu_R^2)}{\pi} 
\left\{c^{{\rm PIM}}_{3 \; gg} \left[\frac{\ln(1-z)}{1-z}\right]_+
+c^{{\rm PIM}}_{2 \; gg} \left[\frac{1}{1-z}\right]_+
+c^{{\rm PIM}}_{1 \; gg}  \delta(1-z) \right\}
\nonumber \\ &&
{}+\frac{\alpha_s^{3}(\mu_R^2)}{\pi} 
\left[A^c_{gg} \,\left[\frac{1}{1-z}\right]_+ 
+T^{c\, {\rm PIM}}_{1\; gg} \, \delta(1-z)\right]\, .
\label{NLOggPIM}
\eeqa
Here the Born term is
\beq
F^{B\; {\rm PIM}}_{gg}=\frac{\beta}{2s}F^{B\; \rm 1PI}_{gg}|_{{\rm PIM}} \, .
\eeq
In addition,
$c^{{\rm PIM}}_{3 \; gg}=4C_A$,
\beq
c^{{\rm PIM}}_{2 \; gg}=-2C_A-2C_A\ln\left(\frac{\mu_F^2}{s}\right)
\equiv T^{{\rm PIM}}_{2 \; gg}-2C_A\ln\left(\frac{\mu_F^2}{s}\right)\, ,
\eeq
\beq
c^{{\rm PIM}}_{1 \; gg}=-\frac{\beta_0}{2}\ln\left(\frac{\mu_F^2}{s}\right) 
+\frac{\beta_0}{2}\ln\left(\frac{\mu_R^2}{s}\right)\, .
\eeq 
Finally,
\beq
T^{{\rm PIM}}_{1 \; gg}=2 T^{\rm 1PI}_{1 \; gg}|_{{\rm PIM}}
+\frac{1}{\alpha_s^2}s\, 
\frac{d^2{\sigma'}^{(1)\; {\rm S+MF} }_{gg}}{dM^2 \, d\cos\theta}
-\frac{1}{\alpha_s^2}
\frac{\beta}{s} \; s^2\, \frac{d^2{\sigma'}^{(1)\; {\rm S+MF} }_{gg}}
{dt_1 \, du_1}|_{{\rm PIM}} \, .
\eeq
Here ${\sigma'}^{(1)\; {\rm S+MF} }_{gg}$ denotes the soft and 
mass factorization 
subtraction terms calculated in Ref.~\cite{KLMV}.
The prime indicates that we drop the overall $\delta(1-z)$ or
$\delta(s_4)$ coefficients from the expressions in Eqs.~(82), (A10), and 
(A11) of Ref.~\cite{KLMV}.

The NNLO soft-plus-virtual corrections in PIM kinematics are
\beqa
&& \hspace{-12mm} 
s\, \frac{d^2{\hat\sigma}^{(2)\; {\rm PIM}}_{gg}}{dM^2 \, d\cos\theta}
=F^{B\; {\rm PIM}}_{gg} \frac{\alpha_s^2(\mu_R^2)}{\pi^2} 
\left\{\frac{1}{2} \left(c^{{\rm PIM}}_{3 \;gg}\right)^2 
\left[\frac{\ln^3(1-z)}{1-z}\right]_+
+\left[\frac{3}{2} c^{{\rm PIM}}_{3 \;gg} \, 
c^{{\rm PIM}}_{2 \;q {\overline q}}
-\frac{\beta_0}{4} c_{3 \;gg} ^{{\rm PIM}} \right] 
\left[\frac{\ln^2(1-z)}{1-z}\right]_+
\right.
\nonumber \\ && \hspace{-5mm}
{}+\left[c^{{\rm PIM}}_{3 \;gg} \, c^{{\rm PIM}}_{1 \;gg}
+\left(c^{{\rm PIM}}_{2 \;gg}\right)^2
-\zeta_2 \, \left(c^{{\rm PIM}}_{3 \;gg}\right)^2
-\frac{\beta_0}{2} T^{{\rm PIM}}_{2 \;gg}
+\frac{\beta_0}{4} c_{3\;gg}^{{\rm PIM}} \ln\left(\frac{\mu_R^2}{s}\right)
+2 C_A K\right]
\left[\frac{\ln(1-z)}{1-z}\right]_+
\nonumber \\ && \hspace{-5mm}
{}+\left[c^{{\rm PIM}}_{2 \;gg} \, c^{{\rm PIM}}_{1 \;gg}
-\zeta_2 \, c^{{\rm PIM}}_{2 \;gg} \, c^{{\rm PIM}}_{3 \;gg}
+\zeta_3 \, \left(c^{{\rm PIM}}_{3 \;gg}\right)^2 
+\frac{\beta_0}{4} c_{2\;gg}^{{\rm PIM}} \ln\left(\frac{\mu_R^2}{s}\right)
+{\cal G}_{gg}^{(2)}\right.
\nonumber \\ && \quad \left.
{}+C_A\frac{\beta_0}{4} \ln^2\left(\frac{\mu_F^2}{s}\right)
-C_A K \ln\left(\frac{\mu_F^2}{s}\right)\right]
\left[\frac{1}{1-z}\right]_+
\nonumber \\ && \hspace{-5mm} \left.
{}+R^{{\rm PIM}}_{gg}  \delta(1-z)\right\} 
\nonumber \\ && \hspace{-10mm}
{}+\frac{\alpha_s^{4}(\mu_R^2)}{\pi^2} 
\left\{\frac{3}{2} c^{{\rm PIM}}_{3 \;gg} \, {A'}^c_{gg}\, 
\left[\frac{\ln^2(1-z)}{1-z}\right]_+
+\left[\left(2 \, c^{{\rm PIM}}_{2 \;gg}-\frac{\beta_0}{2}\right){A'}^c_{gg}
+c^{{\rm PIM}}_{3 \;gg} 
T^{c \, {\rm PIM}}_{1 \;gg}
+{F'}^c_{gg}\right] \left[\frac{\ln(1-z)}{1-z}\right]_+
\right.
\nonumber \\ && \hspace{-5mm}
{}+\left[\left(c^{{\rm PIM}}_{1 \;gg}-
\zeta_2 c^{{\rm PIM}}_{3 \;gg}+\frac{\beta_0}{4}
\ln\left(\frac{\mu_R^2}{s}\right)\right){A'}^c_{gg}
+\left(c^{{\rm PIM}}_{2 \;gg}
-\frac{\beta_0}{2}\right) T^{c \, {\rm PIM}}_{1 \;gg}\right] 
\left[\frac{1}{1-z}\right]_+
\nonumber \\ && \hspace{-5mm} \left.
{}+R^{c\, {\rm PIM}}_{gg} \delta(1-z)\right\} \, ,
\label{NNLOggpim}
\eeqa
with 
\beq
{A'}^c_{gg}=\frac{\beta}{2s} {A}^c_{gg}\, ,  \quad \quad
{F'}^c_{gg}=\frac{\beta}{2s} {F}^c_{gg} \, ,
\eeq
where ${A}^c_{gg}$ and ${F}^c_{gg}$ are the 1PI functions
given in the previous subsection.

The virtual corrections
$R^{{\rm PIM}}_{gg}$, $R^{c\, {\rm PIM}}_{gg}$ are also not fully known.
We keep only certain terms that are determined exactly.
The terms multiplying $\delta(1-z)$ that involve
the factorization and renormalization scales are
\beqa
&& F^{B\; \rm PIM}_{gg} \frac{\alpha_s^2(\mu_R^2)}{\pi^2} 
\left[ \ln^2\left(\frac{\mu_F^2}{m^2}\right)
\left\{\frac{3\beta_0^2}{16}-2 \zeta_2 C_A^2 \right\}
-\frac{3\beta_0^2}{8}\ln\left(\frac{\mu_F^2}{m^2}\right)
\ln\left(\frac{\mu_R^2}{m^2}\right)
+\frac{3 \beta_0^2}{16} \ln^2 \left(\frac{\mu_R^2}{m^2}\right) \right.
\nonumber \\ && \quad \quad \left.
{}+\ln\left(\frac{\mu_F^2}{m^2}\right)
\left\{2 C_A \zeta_2 \left[T^{{\rm PIM}}_{2 \; gg}-2C_A
\ln\left(\frac{m^2}{s}\right)\right] -8 C_A^2 \zeta_3
-2 {\gamma'}^{(2)}_{g/g} \right\}
+\frac{\beta_1}{8} \, \ln\left(\frac{\mu_R^2}{m^2}\right) \right]
\nonumber \\ &&
{}+ \frac{\alpha_s^4(\mu_R^2)}{\pi^2} \left\{\left[2C_A \zeta_2 {A'}^c_{gg}
-\frac{\beta_0}{2}T^{c \, {\rm PIM}}_{1 \; gg}\right] 
\ln\left(\frac{\mu_F^2}{m^2}\right)
+\frac{3\beta_0}{4} T^{c \, {\rm PIM}}_{1 \; gg} \ln\left(\frac{\mu_R^2}{m^2}
\right)\right\} \, .
\eeqa

The terms multiplying $\delta(1-z)$ that arise from inversion and do not 
involve the factorization and renormalization scales are 
\beqa
&& \hspace{-22mm} F^{B\; \rm PIM}_{gg} \frac{\alpha_s^2(\mu_R^2)}{\pi^2} 
\left\{
-\frac{\zeta_2}{2}\, \left[T^{{\rm PIM}}_{2 \;gg}
-2C_A \ln\left(\frac{m^2}{s}\right)\right]^2 
+\frac{1}{4}\zeta_2^2 \,\left(c^{{\rm PIM}}_{3 \;gg}\right)^2 \right.
\nonumber \\ && \quad \quad \left.
+\zeta_3 \, c^{{\rm PIM}}_{3 \;gg} \, \left[T^{{\rm PIM}}_{2 \;gg}
-2C_A \ln\left(\frac{m^2}{s}\right)\right]
-\frac{3}{4}\zeta_4 \, \left(c^{{\rm PIM}}_{3 \;gg}\right)^2 \right\}
\nonumber \\ && \hspace{-22mm}
{}+\frac{\alpha_s^4(\mu_R^2)}{\pi^2} \left\{
\left[\zeta_3 c^{{\rm PIM}}_{3 \;gg}-\zeta_2 
\left(T^{{\rm PIM}}_{2 \;gg}-2C_A\ln\left(\frac{m^2}{s}\right)\right)\right]
{A'}^c_{gg}-\frac{\zeta_2}{2} {F'}^c_{gg} \right\}  \, .
\eeqa

\mysection{Partonic cross sections}

Any difference in the integrated cross sections due to kinematics 
choice arises from uncalculated subleading terms.
At leading order (LO) the partonic threshold
condition is exact and there is no difference between the total cross
sections in the two kinematic schemes. However, beyond LO 
additional soft partons are produced and 
there is a difference when not all terms are known.  The total partonic 
cross section may be expressed in terms of dimensionless scaling functions
$f^{(k,l)}_{ij}$ that depend only on $\eta = s/4m^2 - 1$ \cite{KLMV},
\begin{eqnarray}
\label{scalingfunctions}
\sigma_{ij}(s,m^2,\mu^2) = \frac{\alpha^2_s(\mu)}{m^2}
\sum\limits_{k=0}^{\infty} \,\, \left( 4 \pi \alpha_s(\mu) \right)^k
\sum\limits_{l=0}^k \,\, f^{(k,l)}_{ij}(\eta) \,\,
\ln^l\left(\frac{\mu^2}{m^2}\right) \, .
\end{eqnarray} 
These scaling functions all multiply powers of $\ln(\mu^2/m^2)$ and thus do
not depend on $\mu$ themselves. Here we have set $\mu \equiv \mu_F=\mu_R$.
We work in the $\overline{\rm MS}$ scheme throughout.

Previously, we constructed LL, NLL, and NNLL approximations to 
$f_{ij}^{(k,l)}$ in the $q \overline q$ and $gg$ channels for 
$k \leq 2$, $l \leq k$ \cite{KLMV}. 
We now present the full soft-plus-virtual results for the $f_{ij}^{(2,1)}$ 
and $f_{ij}^{(2,2)}$ scaling functions and the partial results for 
$f_{ij}^{(2,0)}$ that include the soft NNNLL and those virtual terms
calculated in sections 2 and 3.

\begin{figure}[htb] 
\setlength{\epsfxsize=1.0\textwidth}
\setlength{\epsfysize=0.50\textheight}
\centerline{\epsffile{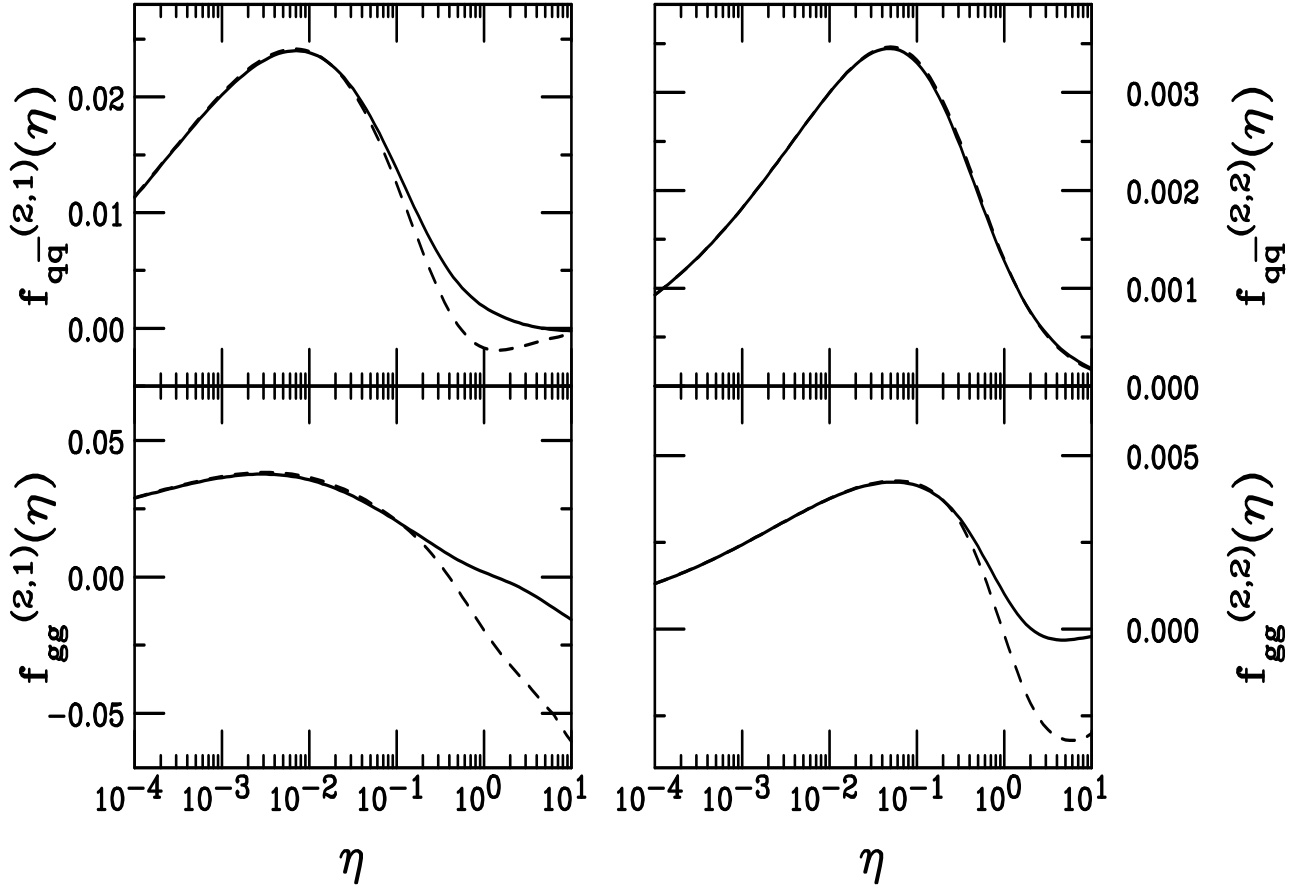}}
\caption[]{The $\overline{\rm MS}$ scheme scaling functions multiplying 
the scale-dependent logarithms,
$f_{ij}^{(2,1)}$ (left-hand side) and $f_{ij}^{(2,2)}$ (right-hand side).
The upper plots are for the $q \overline q$ channel while the 
lower plots are for the $gg$ channel.  The solid
curves are for 1PI kinematics, the dashed for PIM kinematics.
}
\label{fig1} 
\end{figure}

We begin with a comparison of the full soft-plus-virtual 1PI and 
PIM contributions to $f_{ij}^{(2,1)}$ and $f_{ij}^{(2,2)}$, shown in
Fig.~\ref{fig1}.  The upper plots are for the $q \overline
q$ channel.  The left-hand side of
Fig.~\ref{fig1} compares the 1PI and PIM scaling functions for $f_{q \overline
q}^{(2,1)}$.  At low $\eta$,
closer to partonic threshold, the agreement is very good, better than that
obtained at NNLL in Ref.~\cite{KLMV}.  The agreement is also improved 
at large $\eta$.
The right-hand side shows the $f_{q \overline q}^{(2,2)}$ scaling
functions in both kinematics.  The results for  $f_{q \overline q}^{(2,2)}$
remain unchanged from those of Ref.~\cite{KLMV}.  
The agreement between the two kinematics choices is excellent.

The lower plots of Fig.~\ref{fig1} show the corresponding scaling
functions in the $gg$ channel.  The agreement between the two kinematics
choices is somewhat improved at high $\eta$ as compared to previous NNLL 
results \cite{KLMV}.  We note that there is some ambiguity in the way that the
expressions for the $gg$ partonic cross sections 
can be written at threshold.  We have investigated the effect of 
replacing $1 - 2t_1u_1/s^2$ with $(t_1^2 + u_1^2)/s^2$ in 
Eq.~(\ref{acgg}), more consistent with the expressions
in Ref.~\cite{NLOgg}.  These two expressions are equivalent at
threshold, $s_4 = 0$ and $z = 1$, but can differ at large $\eta$.  
Note that $f_{gg}^{(2,2)}$ is not affected by this replacement.
The resulting differences in $f_{gg}^{(2,1)}$ are small, 
appearing only at $\eta > 0.1$ where
the agreement between the scaling functions in the two kinematics
begins to diverge.  The main effect of the second choice is to make
the PIM result for $f_{gg}^{(2,1)}$ more negative at large $\eta$.  
We thus use the expressions as written in the text to be consistent with 
those of Ref.~\cite{KLMV}. 

\begin{figure}[htb] 
\setlength{\epsfxsize=1.0\textwidth}
\setlength{\epsfysize=0.5\textheight}
\centerline{\epsffile{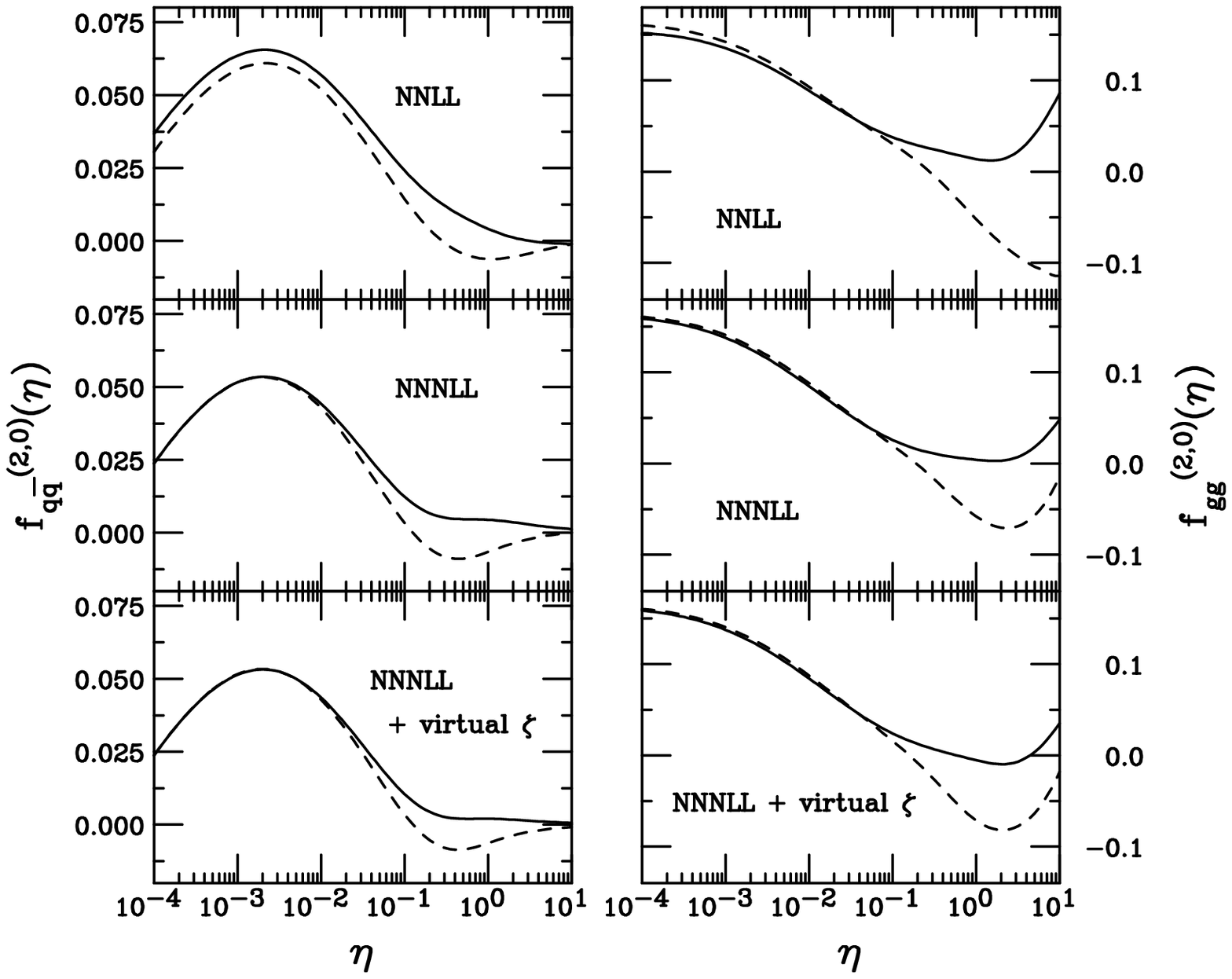}}
\caption[]{The $f_{ij}^{(2,0)}$ scaling functions in the
$\overline{\rm MS}$ scheme.  The left-hand side shows
the results for the $q \overline q$ channel while the right-hand side 
shows the results for the $gg$ channel.  The top plots show the NNLL result 
from Ref.~\protect\cite{KLMV}. The center plots give the results through 
NNNLL and the bottom plots give the results including the virtual
$\zeta$ terms.
The solid curves are for 1PI kinematics, the dashed for PIM kinematics.
}
\label{fig2} 
\end{figure}

We now turn to the $f_{ij}^{(2,0)}$ scaling functions, the most important
contributions at NNLO and independent of $\ln(\mu^2/m^2)$.  We add the
NNNLL terms, i.e. terms proportional to $[1/s_4]_+$ (1PI) and $[1/(1-z)]_+$ 
(PIM), to our previous NNLO-NNLL calculation. We also investigate the
effect of keeping  the virtual $\zeta$ terms  resulting from the 
inversion from moment to momentum space.  To demonstrate the effect of 
adding successive subleading contributions, in Fig.~\ref{fig2} 
we show the NNLL results in
the upper plots, the scaling functions through NNNLL in the middle plots, 
and the results with the NNNLL and virtual $\zeta$ terms in the lower plots.  

We first discuss the results for the $q \overline q$ channel in the
$\overline{\rm MS}$ scheme, shown on the left-hand side of Fig.~\ref{fig2}.
Note that to NNLL, the two kinematics choices give rather different results,
even at low $\eta$.  When the NNNLL contribution is added, both the 1PI and
PIM results are reduced relative to the NNLL over all $\eta$.  The agreement 
between the two kinematics is much improved up to $\eta > 0.01$.  Adding the
virtual $\zeta$ terms resulting from inversion improves the agreement 
between the 1PI
and PIM kinematics further for $0.01 < \eta < 0.1$.  At $\eta > 0.1$, 
the region
where the parton luminosity peaks for $t \overline t$ production at the
Tevatron, the additional virtual $\zeta$ terms provide 
a further small reduction.
With the subleading terms, the 1PI result is smaller than previously but
positive while the PIM result becomes more negative.  However, on the whole,
the subleading terms bring the 1PI and PIM results into better agreement over
all $\eta$.
We note here that the effect of the virtual $\zeta$ terms is numerically 
small, as is also the case for the $gg$ channel
and for the hadronic results for both channels in the next section. 
This small effect is in agreement with the arguments in
Section IIIC of Ref.~\cite{NKtop} concerning resummation prescriptions.
There it was shown that when subleading terms from inversion are
calculated exactly they do not have an unwarrantedly large effect on the
numerical results.

A similar trend is seen for the $gg$ channel on the right-hand side of
Fig.~\ref{fig2}.  The agreement between the NNLL 1PI and PIM scaling functions
at low $\eta$ is significantly better than in the $q \overline q$ channel.
This may perhaps be a consequence of the more complex color structure of 
the $gg$ channel.  Note however the significant
divergence at large $\eta$.  The 1PI NNLL result is large and positive while
the PIM is large and negative.  Again, inclusion of the subleading
contributions improves agreement over all $\eta$.  There is only a small
improvement possible at low $\eta$.  However, the improvement at 
larger $\eta$, $\eta > 0.1$ is notable.  
The 1PI result with soft NNNLL plus virtual $\zeta$ terms 
is reduced by nearly a factor 
of two relative to the NNLL result at $\eta = 10$.  
Likewise, the subleading terms
stop and reverse the downward trend of the PIM scaling functions.  The 1PI $gg$
contribution will still be positive while the PIM will still be negative but 
the difference may not be as large as before.  Using the alternate expression,
$(t_1^2 + u_1^2)/s^2$, in Eq.~(\ref{acgg}) does not significantly change the
results, particularly for 1PI kinematics.  The PIM result becomes slightly more
negative at intermediate $\eta$, $\eta \approx 1$.

Finally we note that if we had kept only the $\zeta$ contributions in the 
$[1/s_4]_+$ and $[1/(1-z)]_+$ 
terms the 1PI and PIM results would not have agreed near threshold. The full 
NNNLL result, given in sections 2 and 3, is required for the result to be
independent of kinematics choice near threshold. 
This agreement also indicates that additional two-loop contributions 
not included in our expressions should be small.

We now turn to our calculations of the hadronic total cross sections and 
transverse momentum distributions.

\mysection{Hadronic total cross sections and $p_T$ distributions}

The inclusive hadronic cross section is obtained by convoluting the
inclusive partonic cross sections with the parton luminosity, $\Phi_{ij}$,
defined as
\begin{eqnarray}
\Phi_{ij}(\tau,\mu_F^2) &=& \tau \,\, \int\limits_{0}^{1}
dx_1\,\, \int\limits_{0}^{1} dx_2\,\, \delta(x_1x_2 - \tau)\,\,
\phi_{i/h_1}(x_1,\mu_F^2)\, \phi_{j/h_2}(x_2,\mu_F^2)\, ,
\label{parlum}
\end{eqnarray}
where $\phi_{i/h}(x,\mu_F^2)$ is the density of partons of flavor $i$ in
hadron $h$ carrying a fraction $x$ of the initial hadron momentum, at
factorization scale $\mu_F$.  Then
\begin{eqnarray}
\sigma_{h_1h_2}(S,m^2) &=& \sum\limits_{i,j = q,{\overline{q}},g} \,\,
\int\limits_{4m^2/S}^{1}\,\frac{d\tau}{\tau}\,\,\Phi_{ij}(\tau,\mu_F^2)\,\,
\sigma_{ij}(\tau S,m^2,\mu_F^2)\,  \label{sigtot} \\
&=& \sum\limits_{i,j = q,{\overline{q}},g} \,\,
\int_{-\infty}^{\log_{10}(S/4m^2-1)} d\log_{10}\eta \, \frac{\eta}{1+\eta} 
\ln(10) \, \Phi_{ij}(\eta,\mu_F^2)\,\,
\sigma_{ij}(\eta,m^2,\mu_F^2)\, \nonumber
\end{eqnarray}
where
\begin{eqnarray} 
\eta = \frac{s}{4 m^2} - 1\, = \, \frac{\tau S}{4 m^2} -1\, ,
\label{eq:etadef}
\end{eqnarray}
and $S$ is the hadronic Mandelstam invariant.
Our investigations in Ref.~\cite{KLMV} showed
that the approximation should hold if the convolution of the parton densities 
is not very sensitive to the high $\eta$ region.

\begin{figure}[htb] 
\setlength{\epsfxsize=1.0\textwidth}
\setlength{\epsfysize=0.3\textheight}
\centerline{\epsffile{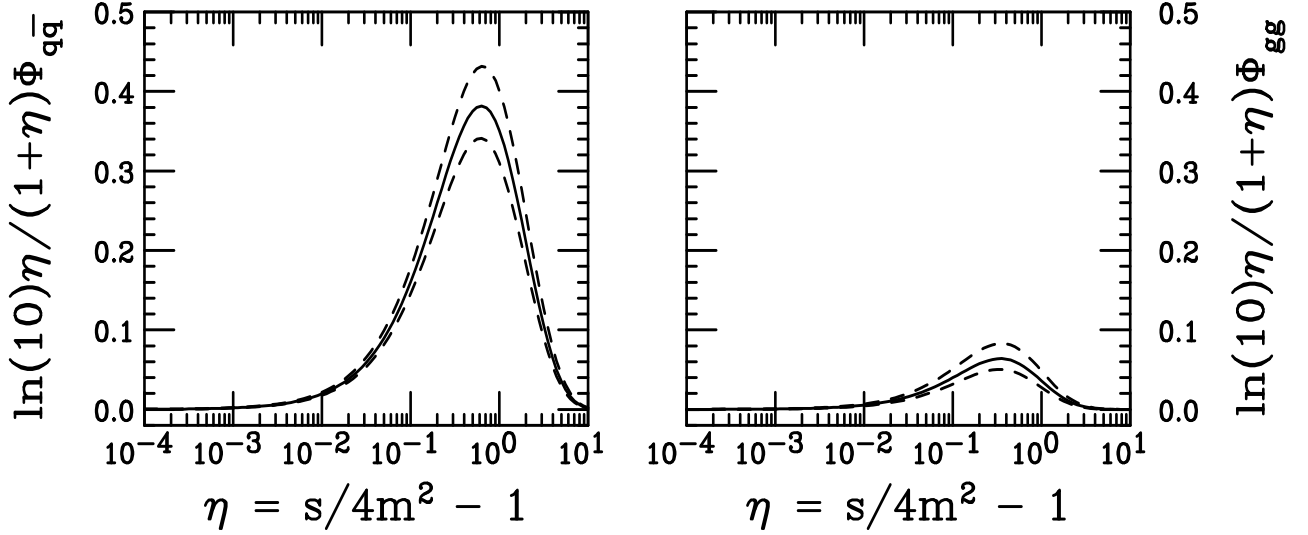}}
\caption[]{The $q \overline q$ (left-hand side) and $gg$ (right-hand side)
parton luminosities in $p  \overline p$ collisions at $\sqrt{S} = 1.96$ TeV.
The solid curves are calculated with $\mu = m = 175$ GeV while the upper 
dashed curves are with $\mu = m/2$ and the lower dashed curves with $\mu = 2m$.
}
\label{tevlum} 
\end{figure}

We use the recent MRST2002 NNLO (approximate) parton
densities \cite{mrst2002} with an NNLO evaluation of $\alpha_s$.  The
parton luminosities, weighted to emphasize 
the most important contributions to the
hadronic cross sections, are shown for $\sqrt{S} = 1.96$ TeV in
Fig.~\ref{tevlum}.  The $q \overline q$ luminosity is nearly 50\% higher than
the CTEQ5M \cite{CTEQ5} $q \overline q$ luminosity used in Ref.~\cite{KLMV}.  
The $gg$ luminosities for the two sets are rather similar.  
The peak of the luminosity
is at $\eta < 1$, but still in a regime where the 1PI and PIM results
differ most.  Fortunately the $gg$ luminosity is small compared to the 
$q \overline q$ luminosity since the differences in the kinematics is largest
in the $gg$ channel.

\begin{figure}[htb] 
\setlength{\epsfxsize=1.0\textwidth}
\setlength{\epsfysize=0.3\textheight}
\centerline{\epsffile{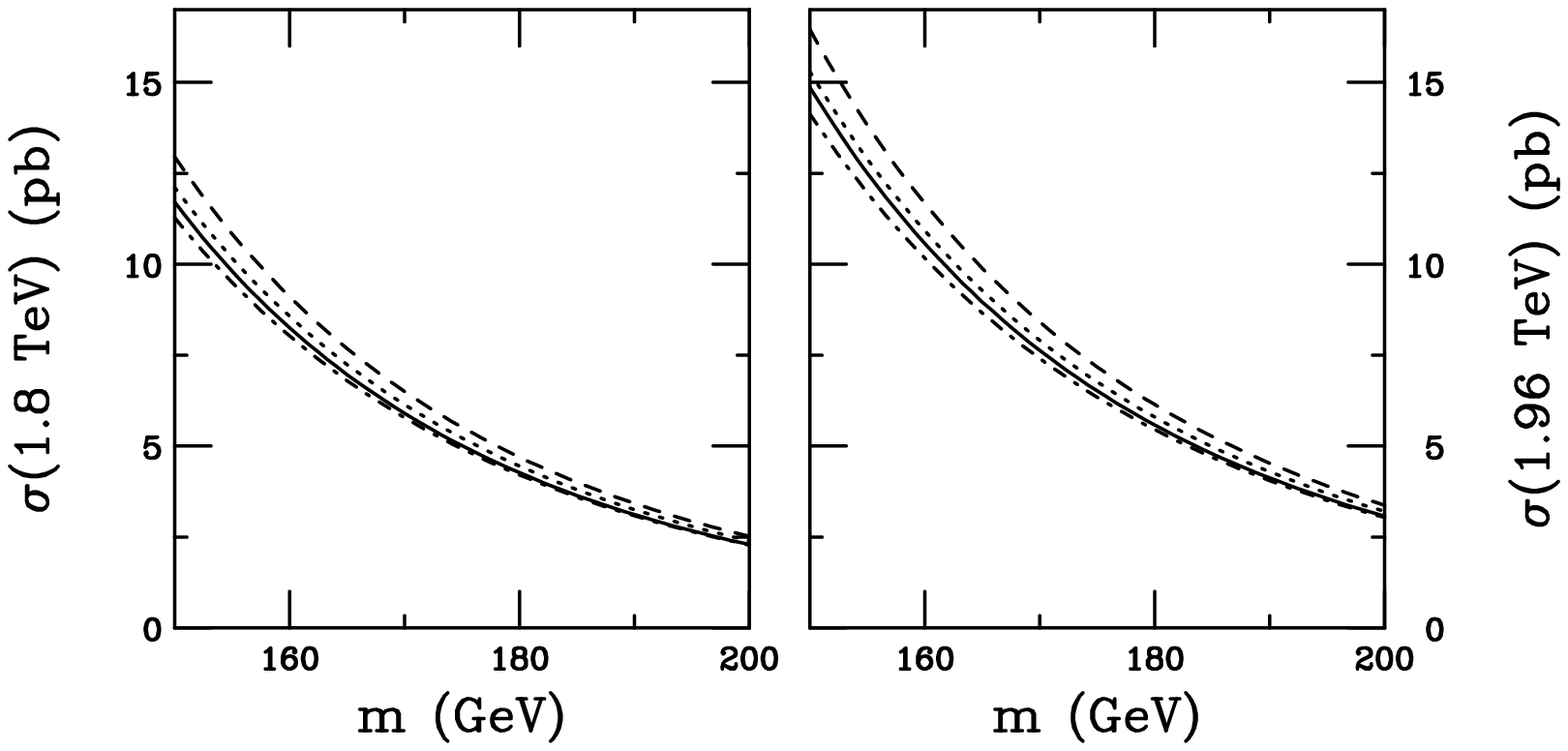}}
\caption[]{The $t \overline t$ total cross sections in $p \overline p$
collisions at $\sqrt{S} = 1.8$ TeV
(left-hand side) and 1.96 TeV (right-hand side) as functions of $m$ for
$\mu = m$.  The NLO (solid), and approximate NNLO 1PI (dashed), 
PIM (dot-dashed) and average (dotted) results are shown.
}
\label{fig4} 
\end{figure}

Our calculations use the exact LO and NLO cross sections with the
soft NNNLL and virtual $\zeta$ corrections and the full soft-plus-virtual
scale-dependent terms at NNLO.  
In addition we multiply the NNLO scaling functions
by a damping factor,
$1/\sqrt{1+\eta}$, as in Ref.~\cite{KLMV}, to lessen the influence
of the large $\eta$ region where the threshold approximation does not
hold so well.

In Fig.~\ref{fig4}, we present the NLO and approximate NNLO $t \overline t$
cross sections at $\sqrt{S} = 1.8$ TeV (left-hand side) and 1.96 TeV
(right-hand side) as functions of top quark mass for $\mu = m$.  The NNLO
results include the soft NNNLL and virtual $\zeta$ terms
in 1PI and PIM kinematics.  
We also show the average of the two kinematics results
which may perhaps be closer to the full NNLO result.  Here the NNLO PIM
cross section is slightly lower than the NLO cross section for all masses 
shown.  In Ref.~\cite{KLMV}, the PIM cross section was a bit higher than 
the NLO.  The reduction
of the PIM $q \overline q$ result, dominant for $p \overline p \rightarrow t
\overline t$, lowers the total PIM cross section.  The NNLO 1PI cross section
remains above the NLO for all $m$ although the NNLO cross  section is not as
large as previously, due to the subleading terms.  The average of the two
kinematics is just above the NLO cross sections for both energies.

\begin{figure}[htb] 
\setlength{\epsfxsize=0.5\textwidth}
\setlength{\epsfysize=0.5\textheight}
\centerline{\epsffile{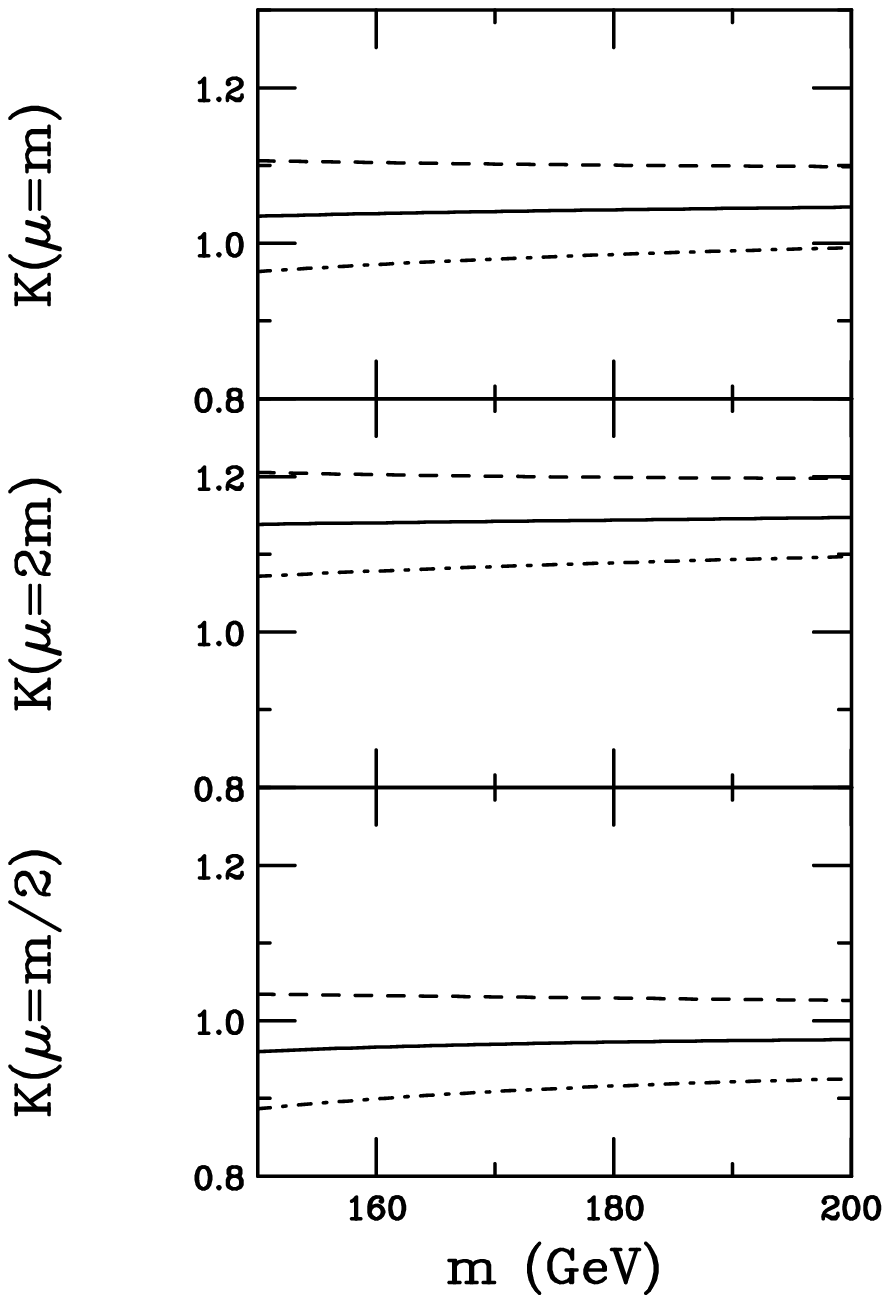}}
\caption[]{The NNLO $K$ factors at $\sqrt{S} = 1.8$ TeV
as functions of top quark mass in $p \overline p$ collisions with $\mu = m$
(upper), $\mu = 2m$ (middle) and $\mu = m/2$ (lower).  The curves show the
ratio of the approximate NNLO 1PI (dashed), PIM (dot-dashed) and 
average (solid) cross sections to the NLO cross section.
}
\label{fig5} 
\end{figure}

Going to higher scales increases all the NNLO corrections so that both
kinematics choices give cross sections larger than the NLO.  On the other hand,
at lower scales, the NNLO cross sections are reduced relative to the NLO.  The
ratio of the NNLO to the NLO cross sections, the $K$ factors, are shown in
Fig.~\ref{fig5} as functions of mass for $\mu = m$ (upper plot), $2m$ (middle
plot and $m/2$ (lower plot) at $\sqrt{S} = 1.8$ TeV.  In keeping with the
results in Fig.~\ref{fig4}, for $\mu = m$ $K<1$ for PIM kinematics, $>1$ for
1PI and for the average.  The $K$ factors are larger
for $\mu = 2m$ and smaller for $\mu = m/2$.  Note also that $K$ is almost
independent of $m$.  We remark that the NLO/LO $K$ factor, while also
essentially mass independent, is typically larger than the NNLO/NLO $K$ 
factors shown here. It is $\sim 1.25$ for $\mu = m$, 1.52 for $\mu = 2m$ and
0.94 for $\mu = m/2$.  Only the last value is similar to that of the NNLO/NLO
average $K$ factor in Fig.~\ref{fig5}.  The small $K$ factors, 
obtained with results calculated with the MRST 
NNLO parton distribution functions at each order, indicate good convergence.
Even though the results are shown at $\sqrt{S} = 1.8$ TeV, the $K$ factors
at $\sqrt{S} = 1.96$ TeV are very similar.

\begin{figure}[htb] 
\setlength{\epsfxsize=1.0\textwidth}
\setlength{\epsfysize=0.3\textheight}
\centerline{\epsffile{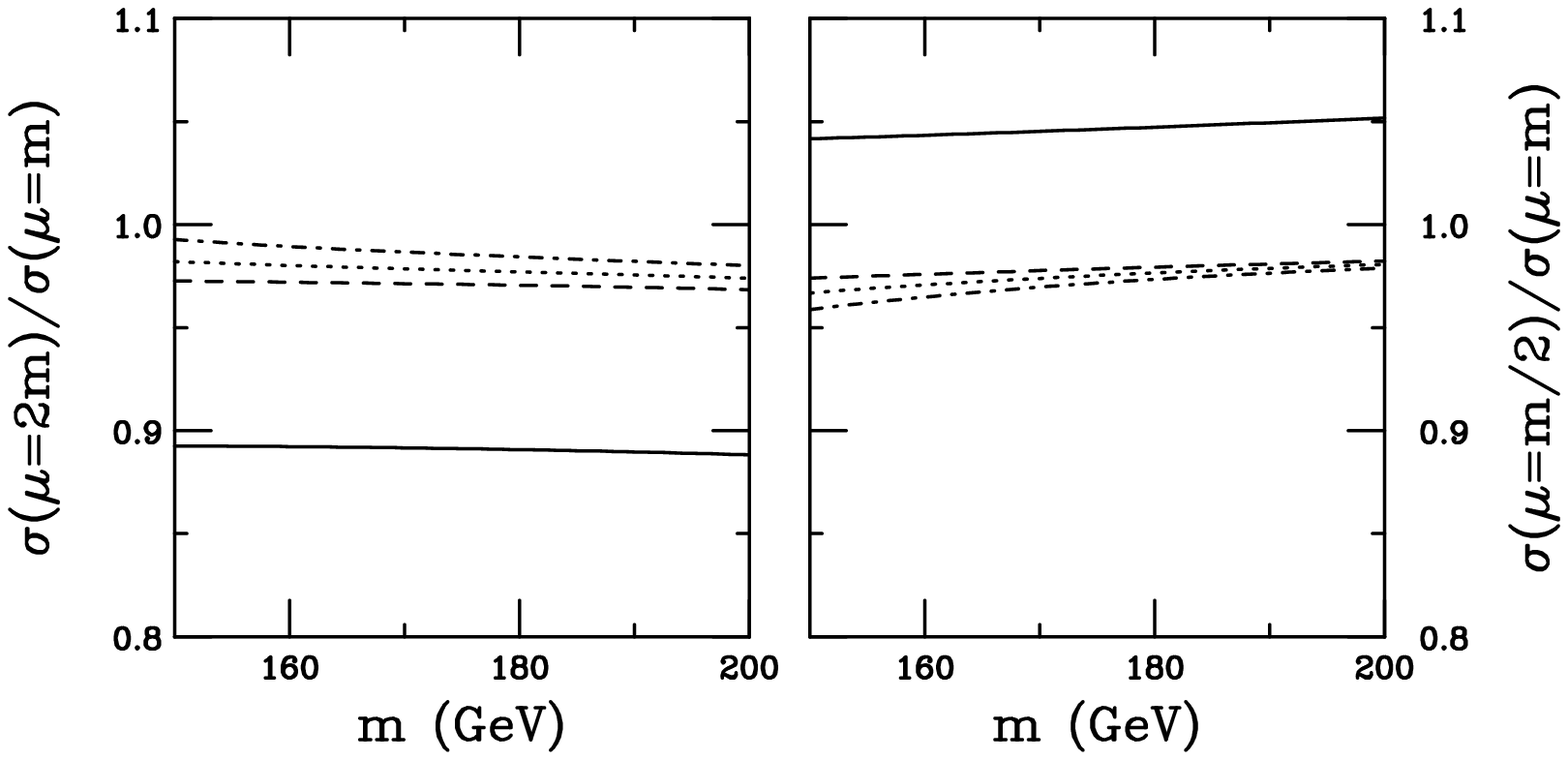}}
\caption[]{The scale dependence of the $t \overline t$ total cross sections 
in $p \overline p$ collisions
at $\sqrt{S} = 1.8$ TeV as a function of top quark mass.  The left-hand side
shows the ratio $(\mu = 2m)$/$(\mu = m)$ while the right-hand side gives the 
ratio for $(\mu = m/2)$/$(\mu = m)$.  The NLO (solid), and approximate
NNLO  1PI (dashed), PIM (dot-dashed) and average (dotted) results are shown.
}
\label{fig6} 
\end{figure}

\begin{figure}[htb] 
\setlength{\epsfxsize=0.75\textwidth}
\setlength{\epsfysize=0.40\textheight}
\centerline{\epsffile{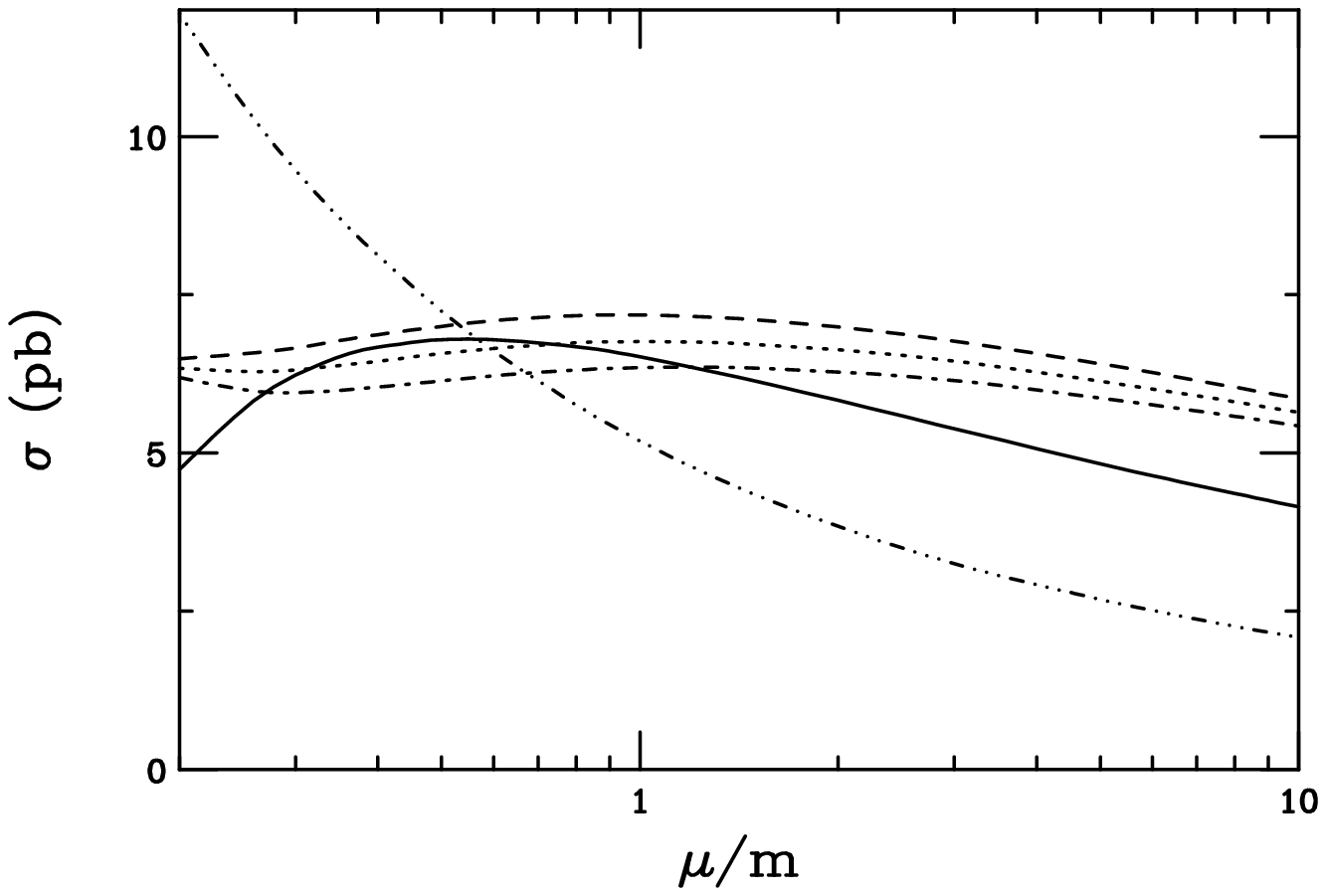}}
\caption[]{The scale dependence of the $t \overline t$ total cross sections 
in $p \overline p$ collisions
at $\sqrt{S} = 1.96$ TeV as a function of $\mu/m$. The LO (dot-dot-dot-dashed),
NLO (solid), and approximate NNLO 1PI (dashed), PIM (dot-dashed) 
and average (dotted) results are shown.
}
\label{fig7} 
\end{figure}

We now examine the scale dependence in Fig.~\ref{fig6} as a function of top
quark mass and in Fig.~\ref{fig7} as a function of 
$\mu/m$ with $m = 175$ GeV.
Figure~\ref{fig6} shows the ratio of the cross sections with $\mu = 2m$ 
to $\mu = m$ on the left-hand side and
the ratio for $\mu = m/2$ to $\mu = m$ on the right-hand side at both NLO 
and NNLO at $\sqrt{S} = 1.8$ TeV.  The ratios
are nearly independent of mass at this energy.  
The scale dependence is reduced
at NNLO relative to NLO.  The NNLO results are very similar for the 
two ratios.  In contrast, the LO scale dependence is much larger, 
$\sigma(\mu =
2m)/\sigma(\mu = m) \approx 0.74$  and $\sigma(\mu =
m/2)/\sigma(\mu = m) \approx 1.4$.  The difference between the scale 
dependence at $\sqrt{S} = 1.8$ TeV and 1.96 TeV is negligible.

We have also calculated the cross sections as functions of $\mu/m$ for
$0.2 < \mu/m < 10$ at $\sqrt{S} = 1.96$ TeV and $m = 175$ GeV
in Fig.~\ref{fig7}.  The NLO cross
section is not as strong a function of $\mu/m$ as the LO cross section.  In
fact, it is seen to rise with $\mu/m$ and then turn over. The
NNLO cross sections, however, exhibit even less dependence on $\mu/m$, 
approaching the independence of scale corresponding to a true physical 
cross section. They change by less than 15\% over the entire range of 
$\mu/m$ considered.  The change in the NNLO cross sections through the
range $m/2 < \mu < 2m$, normally displayed as a measure of uncertainty
from scale variation, is less than 3\%.
Note also that, at
this energy, the absolute difference between the 1PI and PIM cross sections
is also not large.

\begin{table}[htb]
\begin{center}
\begin{tabular}{|c|c|c|c|c|c|c|c|} \hline
\multicolumn{8}{|c|}{$\sigma$ (pb)} \\ \hline 
& & \multicolumn{3}{c}{MRST2002 NNLO} & \multicolumn{3}{|c|}{CTEQ6M} \\ \hline
$\sqrt{S}$ (TeV) & Order & $\mu = m/2$ & $\mu = m$ & $\mu = 2m$ & 
$\mu = m/2$ & $\mu = m$ & $\mu = 2m$ \\ \hline
    & NLO      & 5.24 & 5.01 & 4.46 & 5.27 & 5.06 & 4.51 \\ \hline 
1.8 & NNLO 1PI & 5.40 & 5.52 & 5.36 & 5.43 & 5.58 & 5.43 \\ \hline
    & NNLO PIM & 4.78 & 4.92 & 4.85 & 4.76 & 4.94 & 4.89 \\ \hline  \hline
     & NLO      & 6.79 & 6.52 & 5.83 & 6.79 & 6.54 & 5.85 \\ \hline 
1.96 & NNLO 1PI & 7.00 & 7.17 & 6.99 & 7.01 & 7.21 & 7.04 \\ \hline
     & NNLO PIM & 6.14 & 6.35 & 6.28 & 6.08 & 6.33 & 6.29 \\ \hline 
\end{tabular}
\caption[]{The ${\overline {\rm MS}}$ top quark production cross section in 
$p \overline p$ collisions at the Tevatron ($\sqrt{S} = 1.8$ and 1.96 TeV) 
for $m=175$ GeV.
The NLO results are exact while the approximate NNLO results
include the soft NNNLL corrections and the virtual $\zeta$ terms as well as 
the full soft-plus-virtual scale dependence.}
\end{center}
\end{table}

In Table 1, we give the NLO, NNLO 1PI and NNLO PIM $t \overline t$
total cross sections at $\sqrt{S} = 1.8$ and 1.96 TeV for $p \overline p$
interactions, corresponding to Tevatron Runs I and II.  
The results are presented for $m = 175$ GeV and $\mu = m/2$,
$m$, and $2m$.  We show the results of our calculations
with the MRST2002 NNLO parton densities \cite{mrst2002}
and the three-loop $\alpha_s$.  We compare these with results of
calculations with the CTEQ6M NLO parton densities \cite{cteq6} 
and the two-loop $\alpha_s$.  
The results with the two different sets of parton densities 
are quite similar even 
though the parton densities are evaluated to different orders.  
Note that the NNLO scale dependence is negligible compared to
the NLO scale dependence. The kinematics dependence of the NNLO
cross sections thus remains the largest source of uncertainty.
At $\sqrt{S} = 1.8$ TeV, averaging over the 1PI and PIM NNLO results with 
the two sets of parton distributions at $\mu=m=175$ GeV, 
our best estimate for the cross section is
$5.24 \pm 0.31$ pb where the quoted uncertainty is from the kinematics
dependence. At $\sqrt{S} = 1.96$ TeV our corresponding best estimate is
$6.77 \pm 0.42$ pb.

We note that the cross sections presented in Table 1 are significantly
lower than our previous estimates \cite{NKtop,KLMV} at both NLO and NNLO.
The difference at NLO is solely due to the new sets of parton densities
used here, MRST2002 and CTEQ6M, relative to CTEQ5M in 
Refs.~\cite{NKtop,KLMV}. Our new NLO results are around 3\% lower.
The effect of the new densities on the NNLO corrections is even larger. 
The NNLO-NNLL 1PI corrections are smaller than our previous results 
\cite{NKtop,KLMV} by 14\% for $\mu=m$ and 18\% 
for $\mu=2m$ with the MRST2002 NNLO densities.  Most of this difference
is in the relative values of $\alpha_s$ between the two sets.
In addition, the new subleading terms we have included here further reduce
the magnitude of the NNLO corrections. The combined effect of the new
parton densities and new subleading terms make our new estimates
for the total NNLO $t \overline t$ cross section noticeably smaller.

\begin{figure}[htb] 
\setlength{\epsfxsize=1.0\textwidth}
\setlength{\epsfysize=0.3\textheight}
\centerline{\epsffile{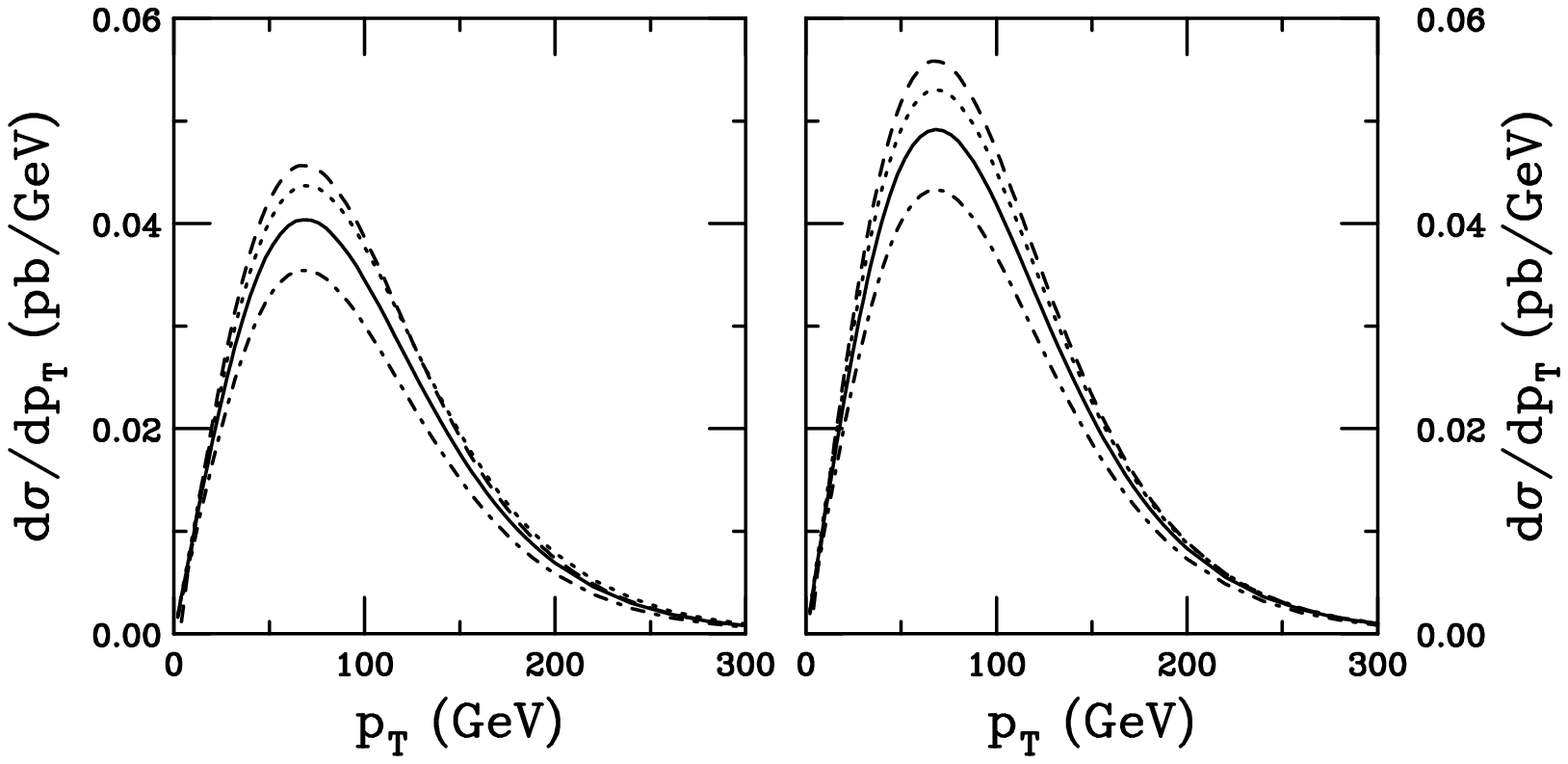}}
\caption[]{The top quark transverse momentum distribution
with $m = 175$ GeV at $\sqrt{S}=1.8$ TeV (left) and 1.96 TeV (right).  
The NLO (solid $\mu=m$, dotted $\mu=m/2$, dot-dashed $\mu=2m$), 
and approximate NNLO ($\mu=m$) 1PI (dashed) results are shown.}
\label{fig8} 
\end{figure}

In Fig.~\ref{fig8} we show the top quark transverse momentum
distributions at $\sqrt{S} = 1.8$ and 1.96 TeV. The NLO and NNLO 1PI
results are shown using the MRST2002 NNLO densities. 
Details of the hadronic calculation of the $p_T$ dependence are given
in Appendix B of Ref. \cite{NKtop}.
At NNLO we observe an enhancement of the NLO distribution
with no significant change in shape. This pattern agrees with 
earlier resummed
results on top transverse momentum and rapidity distributions
\cite{NKtop,NKJS}.

\begin{figure}[htb] 
\setlength{\epsfxsize=1.0\textwidth}
\setlength{\epsfysize=0.3\textheight}
\centerline{\epsffile{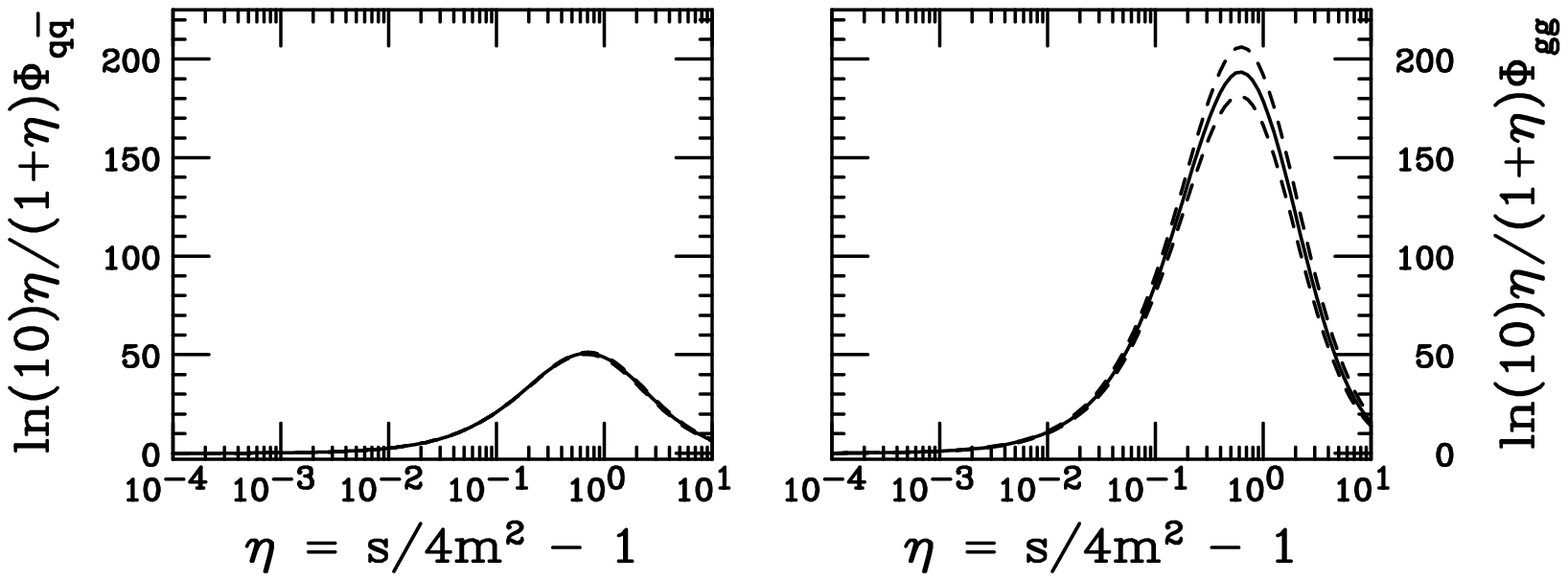}}
\caption[]{The $q \overline q$ (left-hand side) and $gg$ (right-hand side)
parton luminosities in $pp$ collisions at $\sqrt{S} = 14$ TeV.
The solid curves are calculated with $\mu = m = 175$ GeV while the upper 
dashed curves are with $\mu = m/2$ and the lower dashed curves with $\mu = 2m$.
}
\label{lhclum} 
\end{figure}

Finally we discuss top production in $pp$ collisions at the LHC.  
The weighted parton luminosities are shown in Fig.~\ref{lhclum}
for the maximum LHC $pp$ energy, $\sqrt{S} = 14$ TeV.  The $gg$ 
luminosity now dominates the $q \overline q$ by a factor of 4.  
The peak of the luminosity
is still at $\eta \leq 1$ so that this energy is not very far from partonic
threshold.  However, large uncertainties may be expected in the 
$gg$ channel since the difference in the kinematics choice, largest in
this channel, will be emphasized by the high
$gg$ luminosity.

\begin{figure}[htb] 
\setlength{\epsfxsize=0.5\textwidth}
\setlength{\epsfysize=0.3\textheight}
\centerline{\epsffile{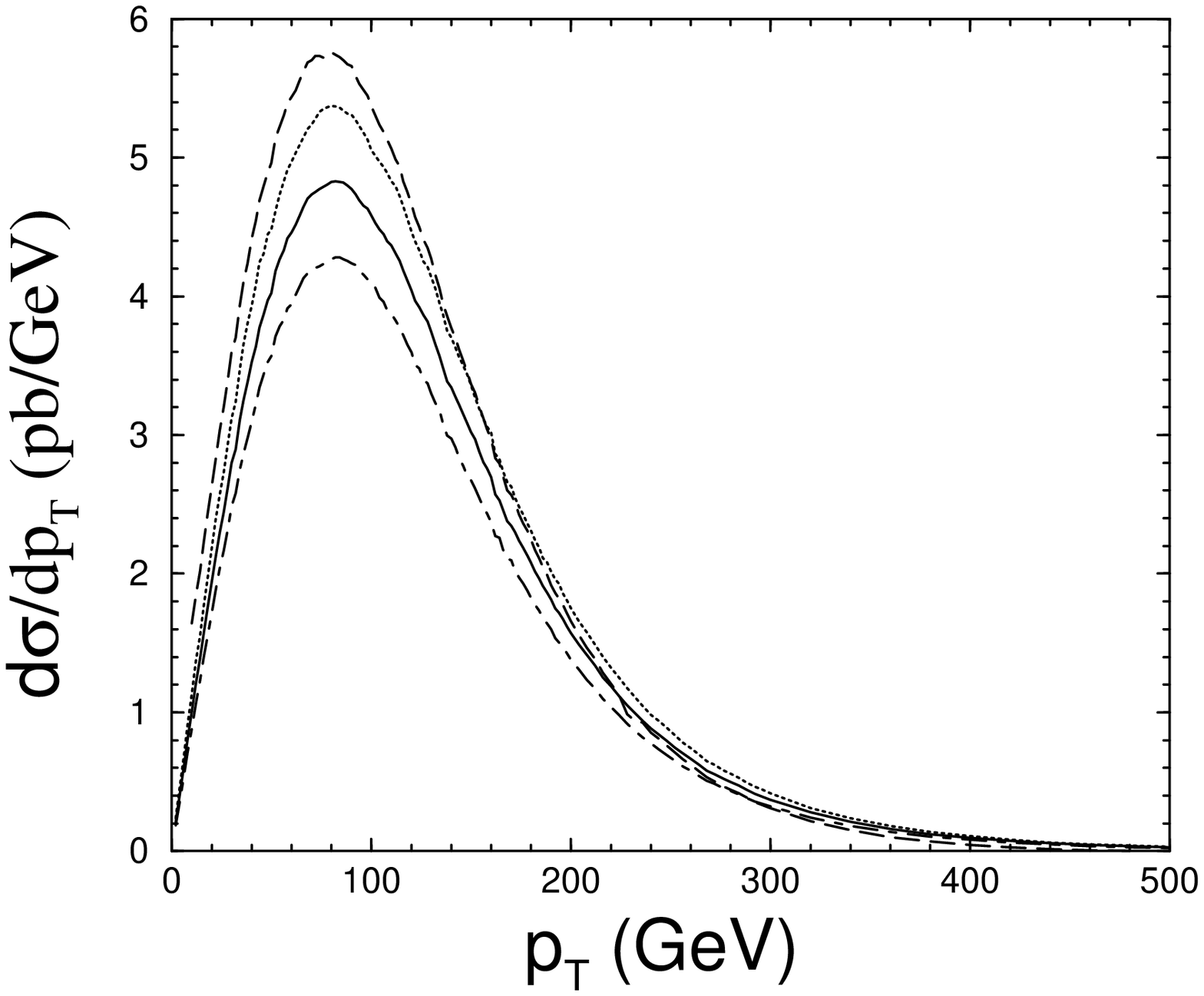}}
\caption[]{The top quark transverse momentum distribution
with $m = 175$ GeV at $\sqrt{S}=14$ TeV.  
The NLO (solid $\mu=m$, dotted $\mu=m/2$, dot-dashed $\mu=2m$), 
and approximate NNLO ($\mu=m$) 1PI (dashed) results are shown.}
\label{fig10} 
\end{figure}

Since the $gg$ contribution dominates at high energy, the difference 
in the results for the two kinematics increases strongly with energy.  
The complex color structure of the $gg$ channel may be better suited to
1PI kinematics and thus this kinematics choice could be more appropriate
in  processes where the $gg$ channel dominates, see Ref.~\cite{KLMVbc} for
discussions of bottom and charm production.  The NNLO 1PI scale dependence
at high energy seems to support such a conclusion.  At $\sqrt{S} = 14$ TeV, 
the NNLO 1PI scale dependence is 4\%, smaller than the 9\% dependence of
the NLO cross section, an acceptable behavior, similar to that at the 
Tevatron.  However, the 
NNLO $gg$ PIM contribution is large and negative.  The $q \overline q$ PIM
contribution is also negative for $\mu \leq m$ albeit much smaller 
than the $gg$ contribution.  The NNLO PIM cross section 
is reduced by nearly a factor of two relative to the NLO.  The scale
dependence is similarly large.  Thus we will only provide NNLO 1PI results
for the LHC.  At $\sqrt{S} = 14$ TeV with $m = 175$ GeV and the MRST2002 NNLO
parton densities, the NLO cross 
section is 808.8 pb for $\mu=m/2$, 794.1 pb for $\mu=m$, 
and 744.4 pb for $\mu=2m$. 
The corresponding NNLO 1PI cross sections are 845.2 pb for $\mu=m/2$, 
872.8 pb for $\mu=m$, and 875.1 pb for $\mu=2m$.
In Fig.~\ref{fig10} we show the NLO and NNLO 1PI top quark $p_T$
distributions at $\sqrt{S} = 14$ TeV.  Here also the NNLO corrections
enhance the NLO result without a change in shape.

\mysection{Conclusions}
In this paper we have calculated soft NNLO corrections
to the total top quark cross section and top transverse momentum distributions
in hadron-hadron collisions. We have added new soft NNNLL terms
and some virtual terms, including all soft-plus-virtual
factorization and renormalization scale terms. 
We have found that these new subleading corrections
greatly diminish the dependence of the cross section on the kinematics
and on the factorization/renormalization scales. 
We have provided numerical results for the total cross section
and top transverse momentum distributions for top quark
production at the Tevatron,
at both Run I and II, and at the LHC.

\mysection*{Acknowledgements}

The research of N.K. has been supported by a Marie Curie Fellowship of 
the European Community programme ``Improving Human Research Potential'' 
under contract number HPMF-CT-2001-01221.
The research of R.V. is supported in part by the 
Division of Nuclear Physics of the Office of High Energy and Nuclear Physics
of the U.S. Department of Energy under Contract No. DE-AC-03-76SF00098.

\end{document}